\documentclass[11pt,a4paper]{article}
\pdfoutput=1
\usepackage{jheppub}
\usepackage{amsmath}
\usepackage{epsfig}
\usepackage{amssymb}
\usepackage{graphics}
\usepackage[active]{srcltx}
\usepackage{epstopdf}
\usepackage{pdfsync}
\usepackage{shuffle}
\usepackage{slashed}

\usepackage{tikz}
\usetikzlibrary{decorations.pathmorphing}
\usetikzlibrary{arrows.meta}

\def\spaa #1{\langle #1\rangle}
\def\spbb #1{[#1]}
\def\spab #1{\langle #1]}
\def\spba #1{[#1 \rangle}
\def\I{\tiny \mbox{I}}
\def\II{\tiny \mbox{II}}

\def\d {\mbox{d}}
\def\tr {\mbox{Tr}}

\title{Wilson Lines and Boundary Operators of BCFW Shifts}
\author[a]{Rijun Huang}
\author[b]{, Qingjun Jin}
\author[b]{and Yi Li}

\affiliation[a]{Institute of Theoretical Physics, School of Physics and Technology, Nanjing Normal University, \\ No.1 Wenyuan Road, Nanjing 210046, P.R.China}
\affiliation[b]{Graduate School of China Academy of Engineering Physics, \\ No. 10 Xibeiwang East Road, Haidian District, Beijing, 100193, P.R.China }

\emailAdd{huang@njnu.edu.cn}
\emailAdd{qjin@gscaep.ac.cn}
\emailAdd{yili@gscaep.ac.cn}

\date{\today}
\abstract{Boundary operators are gauge invariant operators whose form factors correspond to boundary contributions of BCFW shifts. In gauge theory, the boundary operators contain infinite series, which are constrained by gauge symmetry. We compute the boundary operators of all possible BCFW shifts in Yang-Mills theory and QCD, and show that the infinite series can be elegantly organized into Wilson lines, which are natural building blocks for non-local gauge invariant operators. We comment on their connection to jet functions and gauge invariant off-shell amplitudes. We also verify our results by studying various BCFW shifts of four and five-point amplitudes.}

\keywords{Boundary Contribution, Wilson Line, Scattering Amplitude}

\begin{document}
\maketitle \flushbottom

%%%%%%%% Version Information %%%%%%%
% last updated 20221013
% Abstract updated 20221013 by Huang
% Section 1 updated 20221013 by Huang with Jin remark
% Section 2 updated 20221013 by Huang with Jin remark
% Section 3 updated 20221013 by Huang with Jin remark
% Section 4 updated 20221013 by Huang with Jin remark
% Section 5 updated 20221013 by Huang
% Appendix A updated 20221010 by Huang
% Appendix B updated 20221012 by Huang
% Appendix C updated 20221012 by Huang

%%%%%%%%%%%%%%%%
\section{Introduction}
\label{sec:intro}
%%%%%%%%%%%%%%%%%%%

The Wilson line is an important geometric concept in gauge field theory that allows one to compare fields in different space-time points, despite the arbitrary convention of local phases. Because of its intimate relation with gauge invariance, it is widely applied in field theories to construct gauge invariant non-local operators, and set up an ideal framework for analyzing infrared (IR) divergence structures. For example in soft-collinear effective theory (SCET) \cite{Bauer:2000ew,Bauer:2000yr,Bauer:2001ct,Bauer:2001yt},
one can assign Wilson lines to both collinear and soft gauge fields, while the collinear Wilson line is helpful to construct gauge invariant operators with collinear fermions or gluons, and the soft Wilson line encodes the structure of soft interaction \cite{Becher:2014oda}. In perturbation theory, the various Wilson lines defined in SCET are then applied to compute the hard, jet and soft functions (see {\sl e.g.}, \cite{Becher:2014oda,Becher:2018gno} and references therein), describing the factorization properties of a colliding process.

Wilson line also plays a crucial role in many topics of scattering amplitudes. It was used to construct the gauge invariant tree-level multi-gluon amplitudes with some off-shell gluons \cite{Cruz-Santiago:2013vta,Kotko:2014aba,Cruz-Santiago:2015nxa,Bork:2016xfn}, which are defined as matrix elements of Wilson lines. Later, following the idea of Britto-Cachazo-Feng-Witten (BCFW) recursion relation \cite{Britto:2004ap,Britto:2005fq}, the recursive structure of off-shell amplitudes is interpreted as a BCFW-like recursion relation by performing a complex shift of the Wilson line slope instead of the original momentum shift \cite{Kotko:2016qxv}. Furthermore, the Maximal-Helicity-Violating (MHV) vertices or MHV Lagrangian \cite{Ettle:2006bw,Ettle:2007qc,Ettle:2008ey} in Cachazo-Svrcek-Witten (CSW) method \cite{Cachazo:2004kj} are shown to be connected to the straight infinite Wilson line on complex plane \cite{Kotko:2017nkx,Kakkad:2020oim,Kakkad:2021uhv,Kakkad:2021dkw}, in which the Wilson line appears as the field transformation between Yang-Mills theory and MHV Lagrangian. Besides, Wilson line operators also provide an useful bridge connecting the classical double copy and BCJ double copy by replacing color factor in Wilson line with kinematic factor \cite{Alfonsi:2020lub}.

In this paper we show that the Wilson line can be used to construct another type of gauge invariant operator, namely {\sl boundary operator} appearing in the story of BCFW recursion relation. The BCFW recursion builds tree amplitudes recursively from lower-point sub-amplitudes, provided the complex valued amplitude vanishes at large $z$. Otherwise a non-zero boundary contribution would appear and should be computed by other means in order to produce correct amplitude results. After years of development, various methods are proposed to deal with the problem of boundary contributions, either by eliminating them or computing them directly. For instance, by introducing auxiliary fields into the Lagrangian, the boundary contribution in $\lambda\phi^4$ scalar theory could be eliminated \cite{Benincasa:2007xk,Boels:2010mj}. In certain gauge theories and Yukawa theory, by analyzing properties of Feynman diagrams the boundary contributions could also be isolated \cite{Feng:2009ei,Feng:2010ku,Feng:2011twa}. Alternatively, the boundary contribution can be directly computed by multiple-step BCFW-like recursion relations  \cite{Feng:2014pia,Jin:2014qya,Feng:2015qna,Jin:2015pua}, or expressed as roots of amplitudes \cite{Benincasa:2011kn,Benincasa:2011pg,Feng:2011jxa}.

Another very powerful tool to analyze the boundary contributions is the background field method, which can be applied to generic quantum field theories. In \cite{Arkani-Hamed:2008bsc}, background field method was firstly applied to analyzing boundary behavior in gauge and gravity theories. Then it is extended to two derivative gauge and (super)gravity theories \cite{Cheung:2008dn}, as well as multiple-line shift situations \cite{Cheung:2015cba,Hu:2020ddj}. Within the theoretical framework of background method, the {\sl boundary operator} is defined from the operator product  expansion of shifted fields, and one can interpret boundary contribution as form factors (or matrix elements) of boundary operators \cite{Jin:2015pua}. This idea can be reversed, and one uses the knowledge of boundary contribution to compute form factor by suitable construction of Lagrangian that generating the corresponding operator \cite{Huang:2016bmv}.

The Wilson line and boundary contribution of amplitude under BCFW shift are apparently two different physics quantities, however we found that the boundary operator sews them together. In light-cone gauge, the boundary operator of Yang-Mills theory only has a single term \cite{Jin:2015pua}. But in generic gauge, the boundary operator contains infinite series which seems hard to organize. In this paper, a careful examination of these infinite series shows that they can be neatly packed into a Wilson line. Hence the complicated boundary contribution admits a universal Wilson line interpretation. We shall prove that such scenario can also be generalized to QCD, where various combinations of BCFW shifts between gluons and quarks can be defined. The boundary contributions at leading large $z$ order can be organized to a single Wilson line. The lower $z$ order boundary operators may contain more than one infinite sums, which can still be organized into Wilson-line-like structures. Thus these results provide a geometric picture for boundary contributions of amplitudes in the BCFW story.

This paper is organized as follows. In \S\ref{sec:ym}, we briefly review the BCFW recursion relations and boundary operators, and prove that in generic gauge, the infinite series of boundary operators in Yang-Mills theory can be described by a single Wilson line. In \S\ref{sec:qcd}, we generalize the Wilson line formalism to QCD, and work out the QCD boundary operators under various BCFW shifts. Difference between boundary contributions of leading and sub-leading large $z$ is clarified. In \S\ref{sec:example}, some examples are presented to demonstrate the Wilson line formalism. Discussions and conclusions are presented in \S\ref{sec:conclu}. Discussions on convention of shifted polarizations and the color structures of different gauge representations are given in Appendix.

%%%%%%%%%%%%%%%%%%%%%%
\section{The boundary operator and the Wilson line}
\label{sec:ym}
%%%%%%%%%%%%%%%%%%%%%%%

The gluon tree amplitudes in Yang-Mills theory behave as $\mathcal{O}(z^3)$ under the $\langle g^-|g^+]$ BCFW shift, and the corresponding boundary operator contains infinite series in a generic gauge \cite{Jin:2015pua}. In this section we evaluate this boundary operator, and show that it can be expressed in terms of a semi-infinite or infinite Wilson line. In \S\ref{subsec:intro-boundary}, we briefly review the boundary contributions and boundary operators. In \S\ref{subsec:YM}, we construct the boundary operator of $\langle g^-|g^+]$-shift, and by evaluating the corresponding Feynman rules, we show that the boundary operator is proportional to $q\cdot A$ multiplied by a Wilson line. Lastly in \S\ref{subsec:4-gluon}, we compute two-gluon form factor of the boundary operator, and verify that it produces the correct boundary contribution of four-gluon tree amplitude under BCFW shifts.

%%%%%%%%%%%%%%%%%
\subsection{BCFW recursion relations and boundary operators}
\label{subsec:intro-boundary}
%%%%%%%%%%%%%%%%%

The BCFW recursion relation \cite{Britto:2004ap,Britto:2005fq} provides an efficient way to calculate the tree-level scattering amplitude, and it was extremely powerful in the computation of high multiplicity tree amplitudes in gauge and gravity theories, which can be difficult to compute via Feynman diagrams. The basic idea of BCFW recursion is to shift two external momenta by a complex momenta $z q$,
\begin{equation}
k_L\rightarrow k_L-zq~~~,~~~k_R\rightarrow k_R+zq~~~\Rightarrow~~~\mathcal{A}\rightarrow \mathcal{A}(z)~,~
\end{equation}
and construct the amplitude $\mathcal{A}$ from the poles of the shifted amplitude $\mathcal{A}(z)$,
\begin{equation}
\mathcal{A}=\frac{1}{2\pi i}\int_{|z|=\infty}\frac{\mathcal{A}(z)}{z}\d z
-\sum_{i}\text{Res}_{z\rightarrow z_i}\frac{\mathcal{A}(z)}{z}~.~\label{bcfw}
\end{equation}
The first term on the r.h.s. of \eqref{bcfw} represents the residue at $z=\infty$, and it vanishes if $\mathcal{A}(z)\rightarrow 0$ as $z\rightarrow \infty$. Then $\mathcal{A}$ can be expressed by the residues at finite $z_i$, which are subsequently factorized into two tree amplitudes with less number of external legs. However, if $\mathcal{A}(z)\sim  \mathcal{O}(z^k)$ with $k\ge 0$, the residue at $z=\infty$ (the so called {\sl boundary contribution}) does not vanish, which breaks the recursion relation,
\begin{equation}
\mathcal{A}(z)=\sum_{i=0}^k z^i B^{(i)}+\mathcal{O}(\frac{1}{z})~~~\Rightarrow~~~
\mathcal{A}=B^{(0)}
-\sum_{i}\text{Res}_{z\rightarrow z_i}\frac{\mathcal{A}(z)}{z}~.~\label{boundary-contribution}
\end{equation}
$B^{(i)}$ will be called the $\mathcal{O}(z^i)$ boundary contribution.

The large $z$ behavior of BCFW shifts can be studied systematically with the help of background field method \cite{Arkani-Hamed:2008bsc,Jin:2015pua}. Each field $\Phi$ is split into a soft background field which is still labelled by $\Phi$, and a hard field $\Phi^{\Lambda}$ which is regarded as quantum perturbation,
\begin{equation}
\Phi\rightarrow \Phi+\Phi^{\Lambda}~.~
\end{equation}
Since the shifted momenta $k\pm zq$ are much larger than the unshifted momenta, the shifted particles correspond to hard fields, while the unshifted particles correspond to soft fields.

The correlation function of two hard fields and $m$ soft fields can be expressed by \cite{Jin:2015pua},
\begin{equation}
\Bigl\langle \Phi^{\Lambda}_L\Phi^{\Lambda}_R\Phi_1\cdots \Phi_m\Bigr\rangle
=\int D\Phi~\exp\Big(iS[\Phi]\Big)G(z,\Phi)\Phi_1\cdots \Phi_m~,~
\end{equation}
where the $z$-dependence is completely captured by $G(z,\Phi)$, which is the Green's function of two hard fields in the background of soft fields. Let $S_2^{\Lambda}[\Phi^{\Lambda},\Phi]$ be the terms quadratic in $\Phi^{\Lambda}$ from the action $S[\Phi+\Phi^{\Lambda}]$,
\begin{equation}
S_2^{\Lambda}[\Phi^{\Lambda},\Phi]=\frac{1}{2}\int \d^Dx~\Phi^{\Lambda}\mathcal{D}(z,\Phi)\Phi^{\Lambda}~.~
\end{equation}
Then we derive
\begin{equation}
G(z,\Phi)=-i\int D\Phi^{\Lambda}~\exp\Big(iS_2^{\Lambda}[\Phi^{\Lambda},\Phi]\Big)\Phi_L^{\Lambda}\Phi_R^{\Lambda}
~~\sim~~ \mathcal{D}^{-1}(z,\Phi)~.~\label{gz}
\end{equation}
After the LSZ reduction of both hard particles,
\begin{equation}
G(z,\Phi)~\stackrel{\text{LSZ}}{\longrightarrow}~  \mathcal{B}(z,\Phi)=\sum_{i=0}^k z^i \mathcal{B}^{(i)}(\Phi)
+\mathcal{O}(\frac{1}{z})~,~
\end{equation}
in which $\mathcal{B}^{(i)}(\Phi)$ is the $\mathcal{O}(z^i)$ order \textbf{boundary operator}\footnote{We use $B$ to denote boundary contribution, and $\mathcal{B}$ to denote boundary operator. Sometimes they are dressed with a superscript $^{(i)}$ showing the $\mathcal{O}(z^i)$ order, and a subscript showing the BCFW shifts.}, and the corresponding form factor equals to the $\mathcal{O}(z^i)$ order boundary contribution,
\begin{equation}
B^{(i)}=\int \d^Dx\Bigl\langle 0\Bigr|\mathcal{B}^{(i)}\Bigl(\Phi(x)\Bigr)\Bigr|k_1,\ldots,k_m\Bigr\rangle~.~
\end{equation}
%
%But it will not give the right results if the boundary contribution doesn't vanish[reference]. When we consider the boundary contribution, the shifted momenta $k_L+zq$ and $k_R-zq$ are larger than other unshifted momenta. Then the shifted momenta can regarded as the hard field, and the unshifted momenta can regarded as the soft field. So the background field method can be used to analyze the boundary contribution. In the article 1507.00463, the authors use the background field method to get the boundary operators and analyze the $z$ behavior of the boundary contribution. They follow the Wilson's idea and split the field into the high energy part $\Phi^{\Lambda}$ and the low energy part $ \Phi$,

%then the corresponding action can be expanded into the form
%
%\begin{equation}
%S[\Phi+\Phi^{\Lambda}]=S[\Phi]+S_1^{\Lambda}[\Phi^{\Lambda},\Phi]+S_2^{\Lambda}[\Phi^{\Lambda},\Phi]
%+\cdots ,
%\end{equation}
%
%where $S_1^{\Lambda}[\Phi^{\Lambda},\Phi]$ is the terms linear in  $\Phi^{\Lambda}$, and $S_2^{\Lambda}[\Phi^{\Lambda},\Phi]$ is the quadratic term of the field $\Phi^{\Lambda}$, and the $\cdots$ represent the terms with the higher power of  $\Phi^{\Lambda}$. Next, we calculate the  the Green's function of the two hard fields because there are only two shifted momentum. Finally, after doing the LSZ reduction to both hard field, we can get the boundary operator.

%%%%%%%%%%%%%%%%%%%%%%%%%%%%%%%%%%%%%%%%%%%
\subsection{The boundary operator in Yang-Mills theory and Wilson line}
\label{subsec:YM}
%%%%%%%%%%%%%%%%%%%%%%%%%%%%%%%%%%%%%%%%%%%

The boundary operators in models with spin $\le \frac{1}{2}$, such as scalar and Yukawa theories, usually consist of finite number of terms (some examples can be found in \cite{Jin:2015pua}).
%In the story of BCFW recursion relation, boundary contribution is in fact a very general existence. In some theories, for instance scalar or Yukawa theory, the boundary contribution is generated by only a few finite terms, hence a related computation is still under control.
However, we will show that in Yang-Mills and other gauge theories, the boundary operators typically contain infinite series. Fortunately, since the boundary operators are gauge invariant, the infinite series must be strongly constrained by gauge symmetry. This instructs us to connect the boundary operators to Wilson lines, which are perfect building blocks of gauge invariant quantities, and can also be written into infinite sums.
%Since gauge theory is strongly constrained by gauge symmetry, it is not difficult to conjecture that these infinite terms can be determined by gauge invariance, and they could be expressed in a gauge invariant form as a whole. This motives us to think about Wilson line, which is a perfect tool to manage the non-local gauge invariant operators.

The boundary operators of Yang-Mills theory was briefly discussed in \cite{Jin:2015pua}. With the background gauge imposed on hard fields, the inverse propagator $\mathcal{D}$ reads
\begin{equation}
{\cal D}^{\mu a;\nu b}=\eta^{\mu\nu}\delta^{ab}\partial^2-g_s{V}^{\mu a;\nu b}~,~\label{ym-D}
\end{equation}
%
%V^{\mu a;\nu b}+2izg_sf^{abc}\eta^{\mu\nu}q\cdot A^c
in which
\begin{equation}
{V}^{\mu a;\nu b}=- f^{abc}\Bigl(2F^{\mu\nu c}+\eta^{\mu\nu}\big\{\partial_{\alpha},A^{\alpha c}\big\}\Bigr)
-g_sf^{ace}f^{bde}A_{\alpha}^cA^{\alpha d}\eta^{\mu\nu}~.~\label{ym-V}
\end{equation}
We will focus on the $\langle g^-_{L}|g^+_{R}]$-shift, which is the only shift with non-zero boundary contribution. After the LSZ reduction,
\begin{eqnarray}
\mathcal{B}(z)&=&\epsilon_{L-}^{\mu}\epsilon_{R+}^{\nu}\partial^2\left({\cal D}^{-1}_{\mu a;\nu b}-\eta_{\mu\nu}\delta_{ab}\left(\partial^2\right)^{-1}\right)\partial^2~\nonumber\\
&=&\epsilon_{L-}^{\mu}\epsilon_{R+}^{\nu}\Bigl(g_sV_{\mu a;\nu b}-g_s^2V_{\mu a;\alpha c}(\partial^2)^{-1}V^{\alpha c}_{\ \ \ ;\nu b}
+g_s^3V_{\mu a;\alpha c}(\partial^2)^{-1}V^{\alpha c}_{\ \ \ ;\beta d}
(\partial^2)^{-1}V^{\beta d}_{\ \ \ ;\nu b}+\cdots\Bigr)~.~\label{ym-G}
\end{eqnarray}
%
%In the Yang-Mills theory, as calculated in the article 1507.00463, after the LSZ reduction, the authors give the expression as following:
%\begin{equation}
%\begin{aligned}
%&\partial^2\left({\cal D}^{-1}_{\mu a;\nu b}-\eta_{\mu\nu}\delta_{ab}(\partial^2)^{-1}\right)\partial^2
%=V_{\mu a;\nu b}-V_{\mu a;\alpha c}(\partial^2)^{-1}V^{\alpha c}_{\ \ \ ;\nu b}+\cdots\\
%\end{aligned}
%\end{equation}
%In the computation, the background gauge are imposed on the hard fields. And ${\cal D}^{\mu a;\nu b}$ is the effective propagator, $V^{\mu a;\nu b}$ is the term containing the soft field $A$. In the large $z$ limit, and under the $\langle 1 | n ]$ shift, we can know
%\begin{equation}
%\begin{aligned}
%&V^{\mu a;\nu b}(z)= V^{\mu a;\nu b}+ 2izg_sf^{abc}\eta^{\mu\nu}q_{\alpha}A^{\alpha c},\\
%\end{aligned}
%\end{equation}
In \eqref{ym-D}, \eqref{ym-V} and \eqref{ym-G}, the $z$-dependence of these quantities is implicit in $\partial_{\mu}$ and $\epsilon_{L,R}$ as,
\begin{eqnarray}
&&\partial_{\mu}\rightarrow \partial_{\mu}+izq_{\mu}~~~,~~~
\partial^2\rightarrow \partial^2+2izq\cdot\partial~~~,~~~
V^{\mu a;\nu b}\rightarrow V^{\mu a;\nu b}-2izf^{abc}\eta^{\mu\nu}q\cdot A^c~,~\\
&&\epsilon_{L-}^{\mu}\rightarrow \epsilon_{L-}^{\mu}-z\frac{\sqrt{2}k_R^{\mu}}{[LR]}~~~,~~~
\epsilon_{R+}^{\nu}\rightarrow\epsilon_{R+}^{\nu}-z\frac{\sqrt{2}k_L^{\nu}}{\langle LR\rangle}~.~
\end{eqnarray}
More details on the shift of polarization vectors can be found in Appendix \S\ref{appendix:polarization}.

Since $V\sim \mathcal{O}(z),\  (\partial^2)^{-1}\sim \mathcal{O}(z^{-1})$, all terms in the parenthesis of \eqref{ym-G} are of order $\mathcal{O}(z)$. Therefore the leading $\mathcal{O}(z^3)$  order $\mathcal{B}^{(3)}_{\langle g^-_L|g^+_R]}$ contains infinite series\footnote{As discussed in \cite{Arkani-Hamed:2008bsc}, all terms in \eqref{ym-G-2} seem to vanish in the light-cone gauge $q\cdot A^c=0$. But light-cone gauge cannot be imposed on the first term because its momentum satisfies $q\cdot k=0$.},
\begin{equation}
\mathcal{B}^{(3)}_{\langle g^-_L|g^+_R]}=2\Bigl( ig_sf^{a_La_Rc}q\cdot A^c+ig_sf^{a_Le_1c_1}q\cdot A^{c_1} \frac{1}{i q \cdot \partial} ig_sf^{e_1a_Rc_2}q\cdot A^{c_2} + \cdots\Bigr)~,~\label{ym-G-2}
\end{equation}
where we have used
\begin{equation}
\eta_{\mu\nu}\epsilon_{L-}^{\mu}(z)\epsilon_{R+}^{\nu}(z)=-z^2+\mathcal{O}(z)~.~
\end{equation}
For compactness, we will define the following matrix-valued gauge field in adjoint representation\footnote{Similarly, in fundamental representation we define $A_{\mu}=A_{\mu}^aT^a$. Furthermore, $\mathbf{A}_{\mu}=A_{\mu}^a\mathbf{T}^a$ will be used to represent the  matrix-valued gauge field without specifying the representation.},
\begin{equation}
A^{\tiny\mbox{adj}}_{\mu}:=A_{\mu}^aT_A^a~~~,~~~(T_A^a)^{bc}=-if^{abc}~.~
\end{equation}
Then \eqref{ym-G-2} can be written into the following matrix-valued form,
\begin{equation}\label{ym-G-3}
\mathcal{B}^{(3)}_{\langle g^-_L|g^+_R]}
=-2g_sq\cdot A^{\tiny\mbox{adj}}\sum_{j=0}^{\infty}\left( \frac{ig_s}{ q \cdot \partial} q\cdot A^{\tiny\mbox{adj}}\right)^j
:= -2g_sq\cdot A^{\tiny\mbox{adj}}L_q~.~
\end{equation}
One important property of $\mathcal{B}^{(3)}_{\langle g^-_L|g^+_R]}$ is that it is in a sense a gauge invariant operator, {\sl i.e.}, its form factor satisfies the following gauge invariance condition,
\begin{equation}
\int \d^Dx~\Big\langle 0~\Big|~\mathcal{B}^{(3)}_{\langle g^-_L|g^+_R]}(x)~\Big|~k_1,\ldots,k_m\Big\rangle\Bigr|_{\epsilon_i\rightarrow p_i}=0~~~,~~\text{for }i=1,\cdots, m~,~
\end{equation}
because it is the boundary contribution $B^{(3)}$ of a $(m+2)$-point tree amplitude. This means the infinite sum $L_q$ in \eqref{ym-G-3} must be a very special physical quantity. It can be checked that it satisfies the following equation,
\begin{equation}
(q \cdot D) L_q = (q \cdot \partial - i g_s q \cdot A^{\tiny\mbox{adj}} )  L_q=0~.~
\end{equation}
We also observe that the Wilson line
\begin{equation}
\mathbf{W}_n(x) = \bold{P} \exp \left( i g_s \int_{-\infty}^0 \d s~ n \cdot A^a(x+s n)\mathbf{T}^a \right)~,~
\end{equation}
where $\bold{P}$ is the {path-ordering operator}, satisfies a similar equation,
\begin{equation}
 (n \cdot D )\mathbf{W}_n = 0~.~
\end{equation}
This implies that $L_q$ is a Wilson line with $n=q$. In order to verify this conjecture, below we will show that the Feynman rules of $L_q$ are the same as that of $W_q^{\tiny\mbox{adj}}$, which is the Wilson line in the adjoint representation. In other words, $L_q$ and $W_q^{\tiny\mbox{adj}}$ are equivalent in momentum space.
%the boundary operator may have some special relations with the Wilson line. But the expression of the boundary operator and the Wilson line is not similar. However, we can use the Feynman rules of the two different methods to find the relation between the boundary operator and the Wilson line.

%In order to calculate the boundary contribution, we need to calculate the form factor of the boundary operator. As the general process of the form factor calculation, the corresponding Feynman rules should be computed. In the point of the Wilson line, the corresponding Feynman rules should also be calculated to get the final results. If the Feynman rules of the two method can be linked with some special relations, we can conclude that the  $\mathcal{B}_{LO}$ and the Wilson line $W_n$ have the same relations. In the following, we will find the relation.

% At first, we calculate the Feynman rules of the boundary operator and the Wilson line. Then we try to find the relation between the Feynman rules. If the Feynman rules of the two method can be linked with some special relations, we can conclude that the  $\mathcal{B}_{3}$ and the Wilson line $W_n$ have the same relations. In the following, we will use the idea to find the relation.

The Feynman rules for the emission of $m$ gluons from the Wilson line (see {\sl e.g.} \cite{Becher:2014oda}) reads,
% easily get by the traditional method, such as the article 1410.1892. At first, we expand the Wilson line, and use the Fourier transformation. The Wilson line becomes
%\begin{align}
%W_n(x) =& 1+   \int \frac{d^4k}{(2\pi)^4} e^{-i k \cdot x}  (-g_s \frac{n^{\mu}}{n \cdot k} t^a )    \tilde{A}^a_{s\mu}(k)    \notag    \\
%+& \frac{g_s^2}{2}  \Big[  \int \frac{d^4k_1}{(2\pi)^4}  \int \frac{d^4k_2}{(2\pi)^4} ~ e^{-i(k_1+k_2) \cdot x} (\frac{n^{\mu_1}n^{\mu_2}}{n \cdot k_2  ~ n \cdot (k_1+k_2)}) \tilde{A}^{\mu_1}(k_1)  \tilde{A}^{\mu_2}(k_2)  \notag    \\
%&+ (\mu_1 \leftrightarrow  \mu_2,k_1 \leftrightarrow k_2) \Big]   + \cdots ,
%\end{align}
%where we employed the notation $A \equiv A^at^a$. Then the Feynman rules can be easily get.
% For example,  the Feynman rules of the emission of one gluon from a Wilson line is as following
%\begin{equation}
%\label{wilsonfeynman}
%V_{w1} = g_s \frac{n^{\mu}}{n \cdot k} \boldsymbol{T}^a ,
%\end{equation}
% the Feynman rules of the emission of two gluons from a Wilson line can be written as
%\begin{equation}
%V_{w2} = g_s^2 n^{\mu_1}n^{\mu_2} [ \frac{\boldsymbol{T}^{a_1}  \boldsymbol{T}^{a_2} }{n \cdot k_2   ~ n \cdot (k_1+k_2)} +   \frac{\boldsymbol{T}^{a_2}  \boldsymbol{T}^{a_1} }{n \cdot k_1   ~ n \cdot (k_1+k_2)} ] ,
%\end{equation}
%
\begin{equation}
V_{m}^{\tiny\mbox{Wilson}} = g_s^m n^{\mu_1} \cdots n^{\mu_m}  \left( \frac{ \mathbf{T}^{a_1} \cdots \mathbf{T}^{a_m}}{(n \cdot k_m )( n \cdot k_{m-1,m}) \cdots (n \cdot k_{1\cdots m}) }
+ \text{permutations}~\{1, \cdots, m\} \right)~,~\label{wilsonfeynman}
\end{equation}
where we have used the notation $k_{i\cdots j} := k_i +k_{i+1} \cdots + k_j$.
%Here we should pay attention that the symbol $\boldsymbol{T}^a$ of the above equation (\ref{wilsonfeynman}) will be different representation in different situation.

Next let us examine the Feynman rules of the boundary operator. For simplicity we will remove the overall factor $-2$ in \eqref{ym-G-3}. The first term $g_sq\cdot A^{\tiny\mbox{adj}}$ gives the Feynman rule for the emission of a single gluon from $-\frac{1}{2}\mathcal{B}^{(3)}_{\langle g^-_L|g^+_R]}$ as,
\begin{center}
\begin{tikzpicture}
%    \draw [help lines, step=0.5] (-1,-1) grid (8,1);
% Feynman rules gg shift
    \draw [decorate, decoration={coil, segment length=3}] (0,0)--(1,0);
    \draw [very thick] (-1,0.05)--(0,0.05) (-1,-0.05)--(0,-0.05);
    \draw [thick, fill=white] (0,0) circle [radius=0.15];
    \draw [thick] (-0.1,-0.1)--(0.1,0.1) (-0.1,0.1)--(0.1,-0.1);
    \node [above] at (1,0) {$\mu$};
    \node [above] at (-1,0) {$a_R$};
    \node [below] at (-1,0) {$a_L$};
    \node [above] at (0.3,0) {$a$};
%    \node [above] at (-0.5,0.15) {$q$};
%    \draw [->] (-0.8,0.15)--(-0.3,0.15);
    \node [] at (5,0) {$=g_s  T_A^{a} q^\mu$};
\end{tikzpicture}
\end{center}
The second term $g_sq\cdot A^{\tiny\mbox{adj}}\frac{ig_s}{q \cdot \partial} q\cdot A^{\tiny\mbox{adj}}$ gives the Feynman rule for the emission of two gluons from $-\frac{1}{2}\mathcal{B}^{(3)}_{\langle g^-_L|g^+_R]}$ as,
\begin{center}
\begin{tikzpicture}
%    \draw [help lines, step=0.5] (-1,-3) grid (11,-1);
% Feynman rules gg shift
    \draw [decorate, decoration={coil, segment length=3}] (1,-3)--(0,-2)--(1,-1);
    \draw [very thick] (-1,-1.95)--(0,-1.95) (-1,-2.05)--(0,-2.05);
    \draw [thick, fill=white] (0,-2) circle [radius=0.15];
    \draw [thick] (-0.1,-2.1)--(0.1,-1.9) (-0.1,-1.9)--(0.1,-2.1);
    \node [right] at (1,-1) {$a_1,\mu$};
    \node [right] at (1,-3) {$a_2,\nu$};
%    \draw [->] (-0.8,-1.85)--(-0.3,-1.85);
%    \node [above] at (-0.5,-1.85) {$q$};
    \node [above] at (-1,-2) {$a_R$};
    \node [below] at (-1,-2) {$a_L$};
    \draw [->] (0.9,-1.4)--(0.4,-1.9);
    \draw [->] (0.9,-2.6)--(0.4,-2.1);
    \node [] at (1, -1.7) {$k_1$};
    \node [] at (1, -2.3) {$k_2$};
    \node [] at (7.35,-2) {$=g_s^2 (   \frac{q^{\mu_1}q^{\mu_2}}{q \cdot k_2} T_A^{a_1}  T_A^{a_2} + \frac{q^{\mu_1}q^{\mu_2}}{q \cdot k_1} T_A^{a_2}  T_A^{a_1})$};
\end{tikzpicture}
\end{center}
%
%In this case, we should pay attention to the following computation
%
%\begin{eqnarray}
%&& \Big( g_s \eta_{\mu\alpha}q\cdot A(k_1)\Big)\frac{1}{  q\cdot \partial}\Big( g_s \eta^{\alpha}_{~\nu}q\cdot A(k_2)\Big)+\Big(k_1\leftrightarrow k_2\Big)\nonumber\\
%&&=g_s^2\left(\frac{q^{\mu_1}q^{\mu_2}}{q\cdot k_2}A_{\mu_1}(k_1)A_{\mu_2}(k_2)+\frac{q^{\mu_1}q^{\mu_2}}{q\cdot k_1}A_{\mu_2}(k_2)A_{\mu_1}(k_1)\right)~.
%\end{eqnarray}
In general, the Feynman rule for the emission of $m$ gluons from $-\frac{1}{2}\mathcal{B}^{(3)}_{\langle g^-_L|g^+_R]}$ is
\begin{equation}
V_{m}^{\tiny\mbox{operator}} =  g_s^m  q^{\mu_1} \cdots q^{\mu_m}  \left( \frac{ T_A^{a_1}  \cdots T_A^{a_m}}{(q \cdot k_m) (q \cdot k_{m-1,m}) \cdots (q \cdot k_{2 \cdots m}) } + \text{permutations~ of}~\{1, \cdots, m\}\right)~.~\label{boundaryFeyn}
\end{equation}
Terms like $q \cdot k_{1 \cdots i}$ appears in \eqref{boundaryFeyn} because each $\frac{ig_s}{q\cdot \partial}$ acts on all fields on its right.

Comparing (\ref{wilsonfeynman}) with (\ref{boundaryFeyn}), we find
\begin{equation}
V_{m}^{\tiny\mbox{operator}} =  (q \cdot k_{1\cdots m}) V^{\tiny\mbox{Wilson}}_{m}\Bigr|_{n\rightarrow q}~.~
\end{equation}
Therefore the boundary operator and the Wilson line are related by
\begin{equation}
\mathcal{B}^{(3)}_{\langle g^-_L|g^+_R]}=2i q \cdot \partial W_q^{\tiny\mbox{adj}}
=-2g_s q \cdot A^{\tiny\mbox{adj}} W_q^{\tiny\mbox{adj}}~,~
\end{equation}
where we have used $(q\cdot D)W_q^{\text{adj}}=0$ in the last step.

In fact, translating a quantity from position $x$ to $x-\infty q$ by the action of Wilson line $\mathbf{W}_q$ is a standard way to construct non-local operators (see {\sl e.g.} \cite{Becher:2014oda} ). Suppose $\Phi(x)$ is a quantity whose gauge transformation is
\begin{equation}
\Phi(x)~\rightarrow~ \Phi(x)\mathbf{V}^{\dagger}(x)~,~
\end{equation}
then $\Phi(x)\mathbf{W}_q$ transforms as
\begin{equation}
\Phi(x)\mathbf{W}_q~\rightarrow~ \Phi(x)\mathbf{W}_q\mathbf{V}^{\dagger}(-\infty q)~.~
\end{equation}
$\mathbf{V}^{\dagger}(-\infty q)=1$ if we consider gauge functions which vanish at infinity, therefore  $\Phi(x)\mathbf{W}_q$ is a gauge invariant non-local operator.

The gauge transformation of $\mathcal{B}^{(3)}_{\langle g^-_L|g^+_R]}$ is a bit trickier. Let us consider $(q\cdot\mathbf{A})\mathbf{W}_q$ in a generic representation, and it is not invariant under the gauge transformation,
\begin{equation}
(q\cdot \mathbf{A}) \mathbf{W}_q~~~\rightarrow~~~
\left(\mathbf{V}q\cdot \mathbf{A}\mathbf{V}^{\dagger}-\frac{1}{ig_s} \mathbf{V}\left(q\cdot\partial \mathbf{V}^{\dagger}\right)\right)\mathbf{V}\mathbf{W}_q\mathbf{V}^{\dagger}(-\infty q)~.~\label{gluon-not-gauge}
\end{equation}
In fact, as will be shown in \S\ref{subsec:4-gluon}, the $\mathcal{B}^{(3)}_{\langle g^-_L|g^+_R]}\rightarrow 2g$ form factor is not gauge invariant unless the gluon external momenta $k_i$ satisfy $q\cdot (k_1+k_2)=0$.

In order to make the gauge invariance of the operator explicit, let us write it into an infinite Wilson line form as,
\begin{equation}
\mathbf{W}_q^{[-\infty,+\infty]}(x) = \bold{P} \exp \left( i g_s \int_{-\infty}^{+\infty} \d s~ q \cdot A^a(x+s q)\mathbf{T}^a \right)~.~\label{infinite-Wilson}
\end{equation}
Gauge transformation of this Wilson line follows,
\begin{equation}
\mathbf{W}_q^{[-\infty,+\infty]}~~\rightarrow ~~\mathbf{V}(+\infty q)\mathbf{W}_q^{[-\infty,+\infty]}\mathbf{V}^{\dagger}(-\infty q) = \mathbf{W}_q^{[-\infty,+\infty]} ~,~
\end{equation}
where we assume $\mathbf{V}(+\infty q)=1$ and $\mathbf{V}^{\dagger}(-\infty q)=1$. Below let us show that this Wilson line formalism is indeed equivalent to the previously defined semi-infinite Wilson line formalism.

In momentum space, the $[-\infty,+\infty]$ Wilson line becomes,
%Then, we expand the Wilson line, and transform the coordinate space into the momentum space. The Wilson line becomes
%
\begin{eqnarray}
&&\mathbf{W}_q^{[-\infty,+\infty]}(x)  =  1 + i g_s \int_{-\infty}^{+\infty} \d s \int \frac{\d^4k}{(2\pi)^4}~ e^{-ik \cdot (x+sq)} q \cdot \widetilde{A}(k) \nonumber    \\
&&- \frac{g_s^2}{2}  \left(\int_{-\infty}^{+\infty} \d s \int_{-\infty}^s \d t   \int \frac{\d^4k_1}{(2\pi)^4}  \int \frac{\d^4k_2}{(2\pi)^4} ~e^{-ik_1 \cdot (x+tq)}e^{-ik_2 \cdot (x+sq)} q \cdot \widetilde{A}(k_1) q \cdot \widetilde{A}(k_2)
+\left(\begin{array}{cc}
            s \leftrightarrow t \\
            k_1 \leftrightarrow k_2
            \end{array}\right)\right) \nonumber \\
&&+ \mbox{higher~order~terms}~.~
\end{eqnarray}
Changing of integration region from semi-infinite to infinite produces different Feynman rules. Let us firstly consider the integration of $\mathcal{O}(g_s)$ order. Focusing on the $\int \d s$ integration, we get
\begin{eqnarray}
W^{[-\infty,+\infty]}_q(x,g_s)&=& i g_s \int \frac{\d^4k}{(2\pi)^4}~ e^{-ik \cdot x} q \cdot \widetilde{A}(k) \int_{-\infty}^{+\infty} \d s ~ e^{-i(k \cdot q) s}\nonumber\\
&=&  i g_s \int \frac{\d^4k}{(2\pi)^4}~ e^{-ik \cdot x} q \cdot \widetilde{A}(k)\Big(~2 \pi  \delta (k \cdot q)~\Big)~,~
\end{eqnarray}
where the representation of delta function $\int_{-\infty}^{+\infty} \d x ~ e^{-i kx} = 2 \pi  \delta (k)$ has been used. The resulting delta function constrains $q\cdot k=0$, which is exactly the condition we imposed to ensure gauge invariance in the previous semi-infinite case. A similar computation of the $\mathcal{O}(g_s^2)$ order term shows,
\begin{equation}
W_q^{[-\infty,+\infty]}(x,g_s^2)
=\frac{g_s^2}{2}  \int \frac{\d^4k_1}{(2\pi)^4}  \int \frac{\d^4k_2}{(2\pi)^4} ~e^{-ik_1 \cdot x}e^{-ik_2 \cdot x} q \cdot \widetilde{A}(k_1) q \cdot \widetilde{A}(k_2) \frac{ 2 \pi i}{q \cdot k_1}   \delta (q\cdot k_{12}) +
\left(\begin{array}{cc}
s \leftrightarrow t \\ k_1 \leftrightarrow k_2 \end{array}\right)~,
\end{equation}
where we firstly perform the semi-infinite integration over $\int \d t$ then infinite integration over $\int \d s$. Again we see the delta function constrains $q\cdot (k_1+k_2)=0$. The $\mathcal{O}(g_s^m)$ order term can be similarly computed, from which we can deduce the Feynman rule for Wilson line emitting $m$ gluons as\footnote{A similar Feynman rule was obtained in \cite{Kotko:2014aba} but in a more complicated manner.},
\begin{equation}
V_{m}^{\infty\tiny\mbox{Wilson}} = g_s^m  \left( \frac{ q^{\mu_1} \cdots q^{\mu_m}  \mathbf{T}^{a_1} \cdots \mathbf{T}^{a_m}}{(q \cdot k_m )( q \cdot k_{m-1,m}) \cdots (q \cdot k_{2\cdots m}) }
+ \text{permutations}~\{1, \ldots, m\} \right)(2\pi)\delta \big(q \cdot k_{1\cdots m}\big) ~.~\label{wilsonfeynman-infinite}
\end{equation}
Comparing with (\ref{boundaryFeyn}), we find
\begin{equation}
V_{m}^{\infty\tiny\mbox{Wilson}} =  2\pi \delta \big(q \cdot k_{1\cdots m}\big) V^{\tiny\mbox{operator}}_{m}~.~
\end{equation}
Here the boundary operator is directly described by Wilson line from $[-\infty,+\infty]$, which is equivalent to $(q\cdot\mathbf{A})\mathbf{W}_q$, but the gauge invariance is manifest.

%The $\mathcal{O}(g_s)$ order of these three operators are all proportional to $g_sq\cdot \mathbf{A}$. (Add graph). The complete boundary operator can be obtained by translating $g_sq\cdot \mathbf{A}$ from position $x$ to $x-\infty q$ by the action of the Wilson line.
%\begin{equation}
%\mathbf{W}_q\equiv \mathbf{W}(x,-\infty q)
%=\mathbf{P}\exp\Bigl[ig\int_{-\infty}^0 ds q\cdot A^a(x+sq)\mathbf{T}^a \Bigr]
%\end{equation}

%In the SCET theory, the momentum $n$ is ... . In the boundary operator, the momentum $q$ comes from the BCFW shift. In this section, we use the Feynman rules of the boundary operators and the Wilson line to find the relation between them.

%From the article 1507.00463, we can know the following diagram can show the
%What's more, the physical picture is also similar. It can be directly known that there is a line in two method. In the boundary operator, the line connects with the gluon that the momentum is not shifted. In the Wilson line, the gluons that the momentum is not shifted are emitted from the Wilson line. And the momentum of the line in two methods is the same and symboled with $q$. In this section, we get the relation between the boundary operator and the Wilson line by the Feynman rules of the two methods.

%%%%%%%%%%%%%%%%%
\subsection{A four-gluon amplitude example}
\label{subsec:4-gluon}
%%%%%%%%%%%%%%%%%%%%

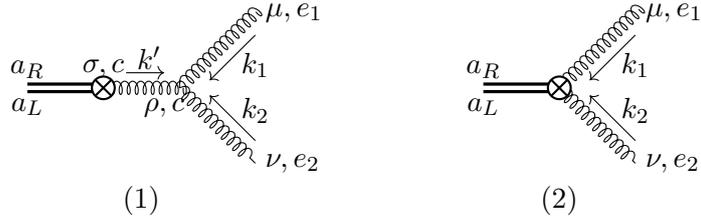
\begin{figure}
\centering
\begin{tikzpicture}
%\draw [help lines, step=0.5] (-1,-1) grid (8,1);
% first diagram
  \draw [very thick] (-1,0.05)--(0,0.05) (-1,-0.05)--(0,-0.05);
  \draw [decorate, decoration={coil, segment length=3}] (0,0)--(1,0) (1,0)--(2,1) (1,0)--(2,-1);
  \draw [thick, fill=white] (0,0) circle [radius=0.15];
  \draw [thick] (-0.1,-0.1)--(0.1,0.1) (-0.1,0.1)--(0.1,-0.1);
  \draw [->] (0.3,0.2)--(0.8,0.2);
  \draw [->] (2,0.7)--(1.4,0.1);
  \draw [->] (2,-0.7)--(1.4,-0.1);
  \node [above] at (-1,0) {$a_R$};
  \node [below] at (-1,0) {$a_L$};
  \node [right] at (2,1) {$\mu, e_1$};
  \node [right] at (2,-1) {$\nu, e_2$};
  \node [below] at (0.8,0) {$\rho, c$};
  \node [above] at (0,0) {$\sigma,c$};
 \node [] at (0.6,0.4) {$k'$};
  \node [] at (2,0.3) {$k_1$};
  \node [] at (2,-0.3) {$k_2$};
% second diagram
  \draw [very thick] (5,0.05)--(6,0.05) (5,-0.05)--(6,-0.05);
  \draw [decorate, decoration={coil, segment length=3}] (6,0)--(7,1) (6,0)--(7,-1);
  \draw [thick, fill=white] (6,0) circle [radius=0.15];
  \draw [thick] (5.9,-0.1)--(6.1,0.1) (5.9,0.1)--(6.1,-0.1);
  \draw [->] (7,0.7)--(6.4,0.1);
  \draw [->] (7,-0.7)--(6.4,-0.1);
  \node [above] at (5,0) {$a_R$};
  \node [below] at (5,0) {$a_L$};
  \node [right] at (7,1) {$\mu, e_1$};
  \node [right] at (7,-1) {$\nu, e_2$};
  \node [] at (7,0.3) {$k_1$};
  \node [] at (7,-0.3) {$k_2$};
% label
  \node [] at (0.5,-1.5) {$(1)$};
  \node [] at (6,-1.5) {$(2)$};
\end{tikzpicture}
\caption{Two Feynman diagrams contributing to the boundary contribution of four-gluon amplitude under $\spab{g_3^-|g_4^+}$-shift. $k'=-k_1-k_2$. The crossed circle with a double-line represents the boundary operator. %$a_L$ and $a_R$ are the color indices of shifted gluons.
}\label{fig:Feynman-4pt-gg}
\end{figure}

In this subsection, we compute the $\mathcal{B}^{(3)}_{\langle g^-_L|g^+_R]}\rightarrow 2g$ tree-level form factor, and verify that it produces the correct $\mathcal{O}(z^3)$ boundary contribution of four-gluon tree amplitude. The Feynman diagrams contributing to the $\mathcal{B}^{(3)}_{\langle g^-_L|g^+_R]}\rightarrow 2g$ tree form factor are shown in Fig.\ref{fig:Feynman-4pt-gg}.
%At first, we use the boundary operator to compute the boundary contribution. In other words, we need to calculate the form factor of the boundary operator. In this case, the corresponding Feynman diagrams is as following.
%
We use Feynman gauge for the three-gluon vertex, and the two diagrams produce the following two terms,
\begin{equation}
\begin{aligned}
B^{a_La_R}
=&2g_s^2f^{a_Lca_R}f^{ca_1a_2} \frac{1}{s_{12}}\Big((\epsilon_1\cdot \epsilon_2)q\cdot (k_1-k_2)+2(\epsilon_2\cdot q)(\epsilon_1\cdot k_2)-2(\epsilon_1\cdot q)(\epsilon_2\cdot k_1)\Big)\\
&-g_s^2\left(f^{a_Lca_1}f^{ca_Ra_2}\frac{(q\cdot \epsilon_1)(q\cdot \epsilon_2)}{q\cdot k_2}+f^{a_Lca_2}f^{ca_Ra_1}\frac{(q\cdot \epsilon_1)(q\cdot \epsilon_2)}{q\cdot k_1}\right)~,\\
\end{aligned}~\label{Bggexp1}
\end{equation}
where we have simplified the expression using $\epsilon_i\cdot k_i=0$.

With above result it can be directly verified that the gauge invariance condition $B^{a_La_R}\Bigr|_{\epsilon_i\rightarrow k_i}=0$ is not satisfied unless the relation $q\cdot (k_1+k_2)=0$ is applied. This means although $\mathcal{B}^{(3)}_{\langle g^-_L|g^+_R]}$ is not a gauge invariant operator by itself, the form factor of $\mathcal{B}^{(3)}_{\langle g^-_L|g^+_R]}$ is gauge invariant with special choice of $q$.

Using $q\cdot (k_1+k_2)=0$ and Bianchi identity for color factors, \eqref{Bggexp1} can be reduced to
\begin{equation}
B^{a_La_R}
=-4g_s^2f^{a_La_Rc}f^{ca_1a_2}\left( \frac{(\epsilon_1\cdot \epsilon_2)(q\cdot k_1)+(\epsilon_2\cdot q)(\epsilon_1\cdot k_2)-(\epsilon_1\cdot q)(\epsilon_2\cdot k_1)}{s_{12}}
+\frac{(q\cdot \epsilon_1)(q\cdot \epsilon_2)}{2q\cdot k_1}\right)~.~\label{Bggexp2}
\end{equation}
In order to compare with the boundary contribution of $\mathcal{A}(1^-,2^+,3^-4^+)$ in spinor-helicity formalism, we replace the polarization vectors by
\begin{equation}
\epsilon_{1\mu}^{-}~\rightarrow~-\frac{\langle1|\gamma_\mu|r_1]}{\sqrt{2}[r_1~1]}~~~,~~~
\epsilon_{2\mu}^{+}~\rightarrow~-\frac{[2|\gamma_\mu|r_2\rangle}{\sqrt{2}\langle r_2~2\rangle}~,~
\end{equation}
and derive\footnote{One may set $|r_1]=|2]$ and $|r_2\rangle=|1\rangle$ to remove most terms.}
\begin{equation}
B^{a_La_R}
=-g_s^2f^{a_La_Rc}f^{ce_1e_2}\frac{\langle1|q|2]^2}{(q\cdot k_1)s_{12}}= -2g_s^2f^{a_3a_4c}f^{ca_1a_2}\frac{\langle14\rangle^4}{\langle12\rangle\langle24\rangle\langle34\rangle\langle41\rangle}~,~\label{Bggexp3}
\end{equation}
where in the last step we have set $|q\rangle=|k_R\rangle=|4\rangle, |q]=|k_L]=|3]$ and $a_L=a_3, a_R=a_4$.

Alternatively, the four-gluon tree amplitude can be expressed by\footnote{The overall factor 4 can be removed by redefining $T^a\rightarrow \sqrt{2}T^a$, as in Gervais-Neveu gauge.}
\begin{equation}
\mathcal{A}(1^-,2^+,3^-,4^+)=4g_s^2\frac{\langle 13\rangle^4}{\langle \sigma_1 \sigma_2\rangle\langle \sigma_2 \sigma_3\rangle\langle \sigma_3 4\rangle\langle 4\sigma_1\rangle}\tr(T^{a_{\sigma_1}}T^{a_{\sigma_2}}T^{a_{\sigma_3}}T^{a_4})
+\text{permutations}~\{\sigma_1,\sigma_2,\sigma_3\}~,
\end{equation}
where $\{\sigma_1,\sigma_2,\sigma_3\}=\{1,2,3\}$. The $\mathcal{O}(z^3)$ boundary contribution under $\langle g_3^-|g_4^+]$-shift is given by
\begin{eqnarray}
B^{(3)}&=&4g_s^2\frac{\langle 14\rangle^4}{\langle 12\rangle\langle 24\rangle\langle 34\rangle\langle 41\rangle}
\Bigl(\tr(T^{a_1}T^{a_2}T^{a_3}T^{a_4})+\tr(T^{a_1}T^{a_4}T^{a_3}T^{a_2})\Bigr)\nonumber\\
&&+4g_s^2\frac{\langle 14\rangle^4}{\langle 12\rangle\langle 24\rangle\langle 43\rangle\langle 41\rangle}
\Bigl(\tr(T^{a_1}T^{a_2}T^{a_4}T^{a_3})+\tr(T^{a_1}T^{a_3}T^{a_4}T^{a_2})\Bigr)\nonumber\\
&=&4g_s^2\tr\big([T^{a_1},T^{a_2}][T^{a_3},T^{a_4}]\big)\frac{\langle 14\rangle^4}{\langle 12\rangle\langle 24\rangle\langle 34\rangle\langle 41\rangle}=-2g_s^2f^{a_1a_2c}f^{a_3a_4c}\frac{\langle 14\rangle^4}{\langle 12\rangle\langle 24\rangle\langle 34\rangle\langle 41\rangle}~.
\end{eqnarray}
This is in agreement with \eqref{Bggexp3}. We have also verified that \eqref{Bggexp2} is consistent with the $\mathcal{O}(z^3)$ boundary contribution of four-gluon amplitude in $D$-dimensional form.

%%%%%%%%%%%%%%%%%%%%%%%%%
\section{Generalizing Wilson line formalism to QCD}
\label{sec:qcd}
%%%%%%%%%%%%%%%%%%%%%%%%%%

Similar as the pure Yang-Mills theory, the boundary operators in generic gauge theories also contain infinite series. We have shown that in Yang-Mills theory all these infinite series can be packed into a term proportional to Wilson line. In this section, we would demonstrate that similar structure also exists in QCD. In \S\ref{subsec:qcd-lagrangian}, we compute the Green's function of hard fields in QCD which is required for the computation of boundary operators. In \S\ref{subsec:qq-shift}, we compute the boundary operator of the fermion-pair shift, and show that it is the same as gluon-gluon shift boundary operator except for an overall color factors. In \S\ref{subsec:g-q-boundary}, we compute the boundary operators of the gluon-quark shifts, and show that they can be written as a Wilson line with quark and gluon field insertions. In \S\ref{subsec:summary-boundary} we summarize the boundary operators we obtained.

%%%%%%%%%%%%%%%%%
\subsection{Boundary operators from QCD Lagrangian}
\label{subsec:qcd-lagrangian}
%%%%%%%%%%%%%%%%%

In order to generalize Wilson line formalism to QCD theory, let us consider the related boundary operators. The QCD theory contains both gluon and quark fields. Following the standard procedure, we split them into hard and soft components as,
\begin{equation}
A_{\mu}^a\rightarrow A_{\Lambda\mu}^{a}+A_{\mu}^a~~~,~~~
\psi\rightarrow \psi_{\Lambda}+\psi~~~,~~~
\bar{\psi}\rightarrow \bar{\psi}_{\Lambda}+\bar{\psi}~.
\end{equation}
Consequently, in QCD Lagrangian we keep only the terms which are quadratic in the hard fields, leading to the following expression,
\begin{equation}
\mathcal{L}_{\Lambda}= \frac{1}{2}A_{\Lambda\mu}^a\left([D^2]^{ab}\eta^{\mu\nu}-2gf^{abc}F^{\mu\nu c}\right)A_{\Lambda\nu}^b
+g_s\bar{\psi}\gamma^{\mu}A_{\Lambda\mu}\psi_{\Lambda}+i\bar{\psi}_{\Lambda}
\gamma^{\mu}D_{\mu}\psi_{\Lambda}
+g_s\bar{\psi}_{\Lambda}\gamma^{\mu}A_{\Lambda\mu}\psi~.
\end{equation}
The quadratic dependence of hard fields can be more explicit when the Lagrangian $\mathcal{L}_{\Lambda}$ is written in the following form
\begin{equation}
\mathcal{L}_{\Lambda}
=\frac{1}{2}\begin{pmatrix}
A_{\Lambda\mu}^a & \bar{\psi}_{\Lambda}& \psi_{\Lambda}^{\top}C
\end{pmatrix}
\mathcal{D}
\begin{pmatrix}
A_{\Lambda\nu}^b    \\
\psi_{\Lambda}\\
C^{-1}\bar{\psi}_{\Lambda}^{\top}\\
\end{pmatrix}~,
\end{equation}
in which $\mathcal{D}$ is a $3\times 3$ matrix independent of hard fields, with explicit definition as,
\begin{equation}
\mathcal{D}=
\begin{pmatrix}
[D^2]^{ab}\eta^{\mu\nu}-2gf^{abc}F^{\mu\nu c}
& g_s\bar{\psi}\gamma^{\mu}T^a
&-g_s\psi^{\top}C\gamma^{\mu}\overline{T}^a\\
g_sT^b\gamma^{\nu}\psi
& i\slashed{D}
&0\\
-g_s\overline{T}^b\gamma^{\nu}C^{-1}\bar{\psi}^{\top}
&0
&-i\overline{\slashed{D}}
\end{pmatrix}~.\label{dmatrix}
\end{equation}
Then coefficient of quadratic product of different hard fields can be read out directly from the corresponding entry of matrix. To derive \eqref{dmatrix}, we have used identities like
\begin{equation}
\bar{\psi}\gamma^{\mu}T^a\psi=-\psi^{\top}(\gamma^{\mu})^{\top}(T^a)^{\top}\bar{\psi}^{\top}
=-\psi^{\top}C\gamma^{\mu}C^{-1}\overline{T}^a\bar{\psi}^{\top}~,
\end{equation}
where $C$ is the charge conjugation matrix, and $\overline{T}^a=-(T^a)^{\top}$, $\overline{\slashed{D}}$ are the $SU(N)$ generator and covariant derivative in the conjugate representation, respectively.

Inspecting the structure of expressions, we can decompose $\mathcal{D}$ into the free part $\mathcal{D}_0$ and the interaction part $\mathbb{V}$ as,
\begin{equation}
\mathcal{D}_0=\begin{pmatrix}
\partial^2\delta^{ab}\eta^{\mu\nu}& 0&0\\
0& i\slashed{\partial}&0\\
0&0&-i\slashed{\partial}
\end{pmatrix}~~~,~~~
\mathbb{V}=
g_s\begin{pmatrix}
V^{\mu a;\nu b}
& \bar{\psi}\gamma^{\mu}T^a
&-\psi^{\top}C\gamma^{\mu}\overline{T}^a\\
T^b\gamma^{\nu}\psi
& \slashed{A}
&0\\
-\overline{T}^b\gamma^{\nu}C^{-1}\bar{\psi}^{\top}
&0
&-\overline{\slashed{A}}
\end{pmatrix}~.\label{dvexp}
\end{equation}
The boundary operators $\mathcal{B}$ of various BCFW shifts are given by corresponding entries of the following matrix,
\begin{equation}
\mathbb{B}=\sum_{j=0}^{\infty}\mathbb{V}\Big(\mathcal{D}_0^{-1}\mathbb{V}\Big)^j~,
\end{equation}
multiplied by external states of hard fields. Each entry is formally an infinite series with increasing orders of $g_s$, and its computation requires the expression of $\mathcal{D}_0^{-1}\mathbb{V}$. From \eqref{dvexp} we obtain
\begin{equation}
\mathcal{D}_0^{-1}\mathbb{V}=g_s(\partial^2)^{-1}
\begin{pmatrix}
V^{\mu a;\nu b}
& \bar{\psi}\gamma^{\mu}T^a
&-\psi^{\top}C\gamma^{\mu}\overline{T}^a\\
-iT^b\slashed{\partial}\gamma^{\nu}\psi
& -i\slashed{\partial}\slashed{A}
&0\\
-i\overline{T}^b\slashed{\partial}\gamma^{\nu}C^{-1}\bar{\psi}^{\top}
&0
&-i\slashed{\partial}\overline{\slashed{A}}
\end{pmatrix}~.
\end{equation}
In the large $z$ limit, $\partial_{\mu}\rightarrow \partial_{\mu}+izq_{\mu}$, and we find
\begin{equation}
\mathcal{D}_0^{-1}\mathbb{V}=\frac{g_s}{2iq\cdot\partial}
\begin{pmatrix}
2q\cdot A^{\tiny\mbox{adj}}
& 0&0\\
T^b\slashed{q}\gamma^{\nu}\psi
& \slashed{q}\slashed{A}
&0\\
\overline{T}^b\slashed{q}\gamma^{\nu}C^{-1}\bar{\psi}^{\top}
&0
&\slashed{q}\overline{\slashed{A}}
\end{pmatrix}+\mathcal{O}(\frac{1}{z})~.\label{dv-behaviour}
\end{equation}
%
%Since $\mathcal{D}_0^{-1}\mathbb{V}$is of the order $\mathcal{O}(z^0)$, $\mathbb{V}(\mathcal{D}_0^{-1}\mathbb{V})^k$ has the same large $z$ behaviour as $\mathbb{V}$.
Note that the leading large $z$ behavior of $(\mathcal{D}^{-1}_{0}\mathbb{V})$ is $\mathcal{O}(z^0)$, thus the multiplication of $(\mathcal{D}^{-1}_{0}\mathbb{V})$ will not increase the $z$-power. The leading large $z$ behavior of boundary operator is determined by $\mathbb{V}$ and external states in the large $z$ limit. We are mostly interested in the boundary operator of leading large $z$, which is related to the leading terms of matrix $\mathbb{B}$. In the large $z$ limit, the $\mathbb{V}_{11}$ entry is $\mathcal{O}(z)$ order, while the others are $\mathcal{O}(z^0)$. As a consequence, for the former case we find the $\mathbb{B}_{11}$ entry is $\mathcal{O}(z)$ order. In fact, the $\mathbb{B}_{11}$ entry corresponds to the shift of two-gluon hard fields, and the infinite series can be compactly packed into a single Wilson line, which is the same as pure Yang-Mills case,
\begin{equation}
\mathbb{B}_{11}=zq\cdot \partial W_q^{\tiny\mbox{adj}}+\mathcal{O}(z^0)~.
\end{equation}
For the latter case, we define
\begin{equation}
\mathbb{B}_{ij}=\mathbb{B}^{0}_{ij}+\mathcal{O}(\frac{1}{z})~~~,~~~\forall (i,j)\ne (1,1)~,
\end{equation}
emphasizing the $\mathcal{O}(z^0)$ order contribution. Since $\mathbb{V}_{11}\sim  \mathcal{O}(z)$, in order to compute $\mathbb{B}^{0}_{12}$ and $\mathbb{B}^{0}_{13}$, we have to expand the $(1,2)$ and $(1,3)$ entries  of \eqref{dv-behaviour} to $\mathcal{O}(\frac{1}{z})$, which would make computation complicated. On the other hand, \eqref{dv-behaviour} is sufficient to compute all $\mathbb{B}^{0}_{2i}$ and $\mathbb{B}^{0}_{3i}$ entries at $\mathcal{O}(z^0)$ level, and since $\mathbb{B}$ is formally a Hermitian matrix, the $\mathbb{B}^{0}_{12}$ and $\mathbb{B}^{0}_{13}$ entries can be obtained from the conjugate of $\mathbb{B}^{0}_{21}$ and $\mathbb{B}^{0}_{31}$. Thus by computing $\mathbb{B}^{0}_{22}, \mathbb{B}^{0}_{33}, \mathbb{B}^{0}_{21}, \mathbb{B}^{0}_{31}$, all boundary operators of various BCFW shifts can be expressed as infinite series of $\mathbb{V}(\mathcal{D}_0^{-1}\mathbb{V})^k$. Similar to $\mathbb{B}_{11}$ case, in the following we would show that for any kind of BCFW shifts, the infinite series can be packed into Wilson line, {\sl i.e.}, boundary contributions of various BCFW shifts in QCD theory possess a Wilson line formalism interpretation.

%%%%%%%%%%%%%%%%%
\subsection{Wilson line for the fermion-pair shift}
\label{subsec:qq-shift}
%%%%%%%%%%%%%%%%%

%%%%%%%%%%%%%%%%%%%%%%%%%%%%%%%%%%%%%%%%%%%
\subsubsection*{The $\langle \psi|\psi]$ and $\langle \bar{\psi}|\bar{\psi}]$-shifts}
%%%%%%%%%%%%%%%%%%%%%%%%%%%%%%%%%%%%%%%%%%%

Let us start with $\mathbb{B}^{0}_{32}$, which corresponds to the $\langle \psi_L|\psi_R]$-shift. From \eqref{dvexp}, $\mathbb{V}_{32}=0$, and it is easy to verify that the $\mathcal{O}(z^0)$ order of $\Bigl[\mathbb{V}(\mathcal{D}_0^{-1}\mathbb{V})^k\Bigr]_{32}$ also vanishes. Therefore, $\mathbb{B}^{0}_{32}=0$, and consequently $\mathbb{B}_{32}=\mathcal{O}(\frac{1}{z})$. Since the external state $|R]\rightarrow |R]+z |L]$\footnote{The spinor with momentum $k_i$ is usually denoted as $|k_i\rangle$ or $|k_i]$, but we also use the notation $|i\rangle, |i]$ for simplicity.}, the $\langle \psi_L|\psi_R]$-shift is at most $\mathcal{O}(z^0)$. More careful examination will find the shift to be a good BCFW shift without boundary contribution, but we will not go into the details. Similarly, the $\langle \bar{\psi}|\bar{\psi}]$-shift, which corresponds to $\mathbb{B}_{23}$,  is also a good BCFW shift\footnote{The $\langle \psi|\psi]$ and $\langle \bar{\psi}|\bar{\psi}]$ may not be good BCFW shifts in generic theories, for example in models with Yukawa couplings.}.

%%%%%%%%%%%%%%%%%%%%%%%%%%%%%%%%%%%%%%%%%%%
\subsubsection*{The $\langle \bar{\psi}|\psi]$ and $\langle \psi|\bar{\psi}]$-shifts}
%%%%%%%%%%%%%%%%%%%%%%%%%%%%%%%%%%%%%%%%%%%

These two shifts correspond to $\mathbb{B}^{0}_{22}$ and $\mathbb{B}^{0}_{33}$, and their expressions are
\begin{equation}
\mathbb{B}^{0}_{22}=\sum_{j=0}^{\infty}g_s\slashed{A}
\left(\frac{ig_s}{2q\cdot\partial}\slashed{q}\slashed{A}\right)^j~~~,~~~
\mathbb{B}^{0}_{33}=-\sum_{j=0}^{\infty}g_s\overline{\slashed{A}}
\left(\frac{ig_s}{2q\cdot\partial}\slashed{q}\overline{\slashed{A}}\right)^j~.~
\end{equation}
The $\mathbb{B}^{0}_{22}$ is related to $\langle \bar{\psi}_L|\psi_R]$-shift, and its corresponding external states of hard fields is shifted as
\begin{equation}
\spab{L|\bullet|R}\to -z^2\spab{R|\bullet|L}+\mathcal{O}(z)~.
\end{equation}
When including the external states, the leading large $z$ behavior is $\mathcal{O}(z^2)$, and the leading boundary operator is
\begin{equation}
\mathcal{B}^{(2)}_{\langle \bar{\psi}_L|\psi_R]}=-\langle R|\mathbb{B}^{0}_{22}|L]
=-g_sq\cdot A W_q~,
\end{equation}
where in the derivation we have taken the advantage of $\slashed{q}=|L]\langle R|+|R\rangle[L|$ and $\slashed{X}\slashed{q}=2q\cdot X-\slashed{q}\slashed{X}$ to obtain
\begin{equation}
\langle  R|\slashed{X}\slashed{q}\cdots \slashed{X}\slashed{q}
=\langle  q|\slashed{X}\slashed{q}\cdots \slashed{X}\slashed{q}
=\langle  R|(2q\cdot X)\cdots (2q\cdot X)~.~\label{gamma-change}
\end{equation}
Similarly the  $\mathbb{B}^{0}_{33}$ is related to  $\langle \psi_L|\bar{\psi}_R]$-shift, and the corresponding external states of hard fields $\spba{L|\bullet|R}$ remains $z$-independent. So the leading large $z$ behavior of this shift is $\mathcal{O}(z^0)$, and the boundary operator is
\begin{equation}
\mathcal{B}^{(0)}_{\langle \psi_L|\bar{\psi}_R]}=[L|\mathbb{B}^{0}_{33}|R\rangle
=-g_sq\cdot \overline{A}\ \overline{W}_q~.
\end{equation}
Comparing the Wilson line formalism of $\spab{g^-_{L}|g^+_R}, \langle \bar{\psi}_L|\psi_R]$ and $\langle \psi_L|\bar{\psi}_R]$-shifts,
\begin{equation}\label{boundary-proportional}
\mathcal{B}^{(3)}_{\langle g^-_L|g^+_R]}=g_sq\cdot A^{\tiny\mbox{adj}}W_q^{\tiny\mbox{adj}}~~~,~~~
\mathcal{B}^{(2)}_{\langle \bar{\psi}_L|\psi_R]}=-g_sq\cdot A W_q~~~,~~~
\mathcal{B}^{(0)}_{\langle \psi_L|\bar{\psi}_R]}=-g_s q\cdot \overline{A}\ \overline{W}_q~,
\end{equation}
we observe that the leading large $z$ contributions of them are proportional to each other except for their $SU(N)$ representations,
\begin{equation}
\mathcal{B}^{(3)}_{\langle g^-_L|g^+_R]}~~~,~~~ \mathcal{B}^{(2)}_{\langle \bar{\psi}_L|\psi_R]}~~~,~~~
\mathcal{B}^{(0)}_{\langle \psi_L|\bar{\psi}_R]}~~\propto ~~g_sq\cdot \mathbf{A}\mathbf{W}_q~,
\end{equation}
where $\mathbf{A}$ and $\mathbf{W}_q$ stands for gauge field and Wilson line in a generic representation. As will be discussed in appendix \S\ref{appendix:rep}, these operators can be simplified using the condition $q\cdot(k_1+\cdots+k_n)=0$, leading to
\begin{equation}
g_sq\cdot \mathbf{A}~\mathbf{W}_q= g_sq\cdot A^a
\Bigl[W_q^{\tiny\mbox{adj}}\Bigr]^{ab}\mathbf{T}^b~.
\end{equation}
Diagrammatically, it means that as far as the states that emitting from Wilson line operators are all gluons, the Feynman rules or Feynman diagrams of different BCFW shifts are equivalent to each other and care nothing about the original hard fields.

\begin{figure}
\centering
\begin{tikzpicture}
%  \draw [help lines, step=0.5] (0,-0.5) grid (12,2);
% first figure
  \draw [very thick] (0,0)--(3,0);
  \draw [decorate, decoration={coil, segment length=3.2}] (1.5,0)--(1.5,1.5);
  \draw [thick, fill=white] (1.5,0) circle [radius=0.15];
  \draw [thick] (1.4,-0.1)--(1.6,0.1) (1.4,0.1)--(1.6,-0.1);
  \node [] at (1.5,-1) {$\spab{g^-_L|g^+_R}$};
  \node [below] at (1.5,0) {{\footnotesize $g_s q^\mu$}};
  \node [above] at (1.5,1.5) {{\footnotesize $(A^{\tiny\mbox{adj}})^{\mu}$}};
% second figure
  \draw [very thick, dashed, ->] (7.5,0)--(6.75,0);
  \draw [very thick, dashed, ->] (6.75,0)--(5.25,0);
  \draw [very thick, dashed] (5.25,0)--(4.5,0);
  \draw [decorate, decoration={coil, segment length=3.2}] (6,0)--(6,1.5);
  \draw [thick, fill=white] (6,0) circle [radius=0.15];
  \draw [thick] (5.9,-0.1)--(6.1,0.1) (5.9,0.1)--(6.1,-0.1);
  \node [] at (6,-1) {$\spab{\bar{\psi}_L|\psi_R}$};
  \node [below] at (6,0) {{\footnotesize $-g_s q^\mu$}};
  \node [above] at (6,1.5) {{\footnotesize $A^{\mu}$}};
% third figure
  \draw [very thick, dashed, ->] (9,0)--(9.75,0);
  \draw [very thick, dashed, ->] (9.75,0)--(11.25,0);
  \draw [very thick, dashed] (11.25,0)--(12,0);
  \draw [decorate, decoration={coil, segment length=3.2}] (10.5,0)--(10.5,1.5);
  \draw [thick, fill=white] (10.5,0) circle [radius=0.15];
  \draw [thick] (10.4,-0.1)--(10.6,0.1) (10.4,0.1)--(10.6,-0.1);
  \node [] at (10.5,-1) {$\spab{\psi_L|\bar{\psi}_R}$};
  \node [below] at (10.5,0) {{\footnotesize $g_s q^\mu$}};
  \node [above] at (10.5,1.5) {{\footnotesize $\overline{A}^{\mu}$}};
\end{tikzpicture}
\caption{At the leading $g_s$ order of boundary operators under $\spab{g^-_L|g^+_R}$, $\spab{\bar{\psi}_L|\psi_R}$ and  $\spab{\psi_L|\bar{\psi}_R}$, the vertices are all represented by a operator emitting a gluon. Despite of color factors, three Feynman rules are the same, showing irrelevance to the actual shifted particles, represented by solid line for gluon and dashed line for fermion.}\label{fig:proportion}
\end{figure}
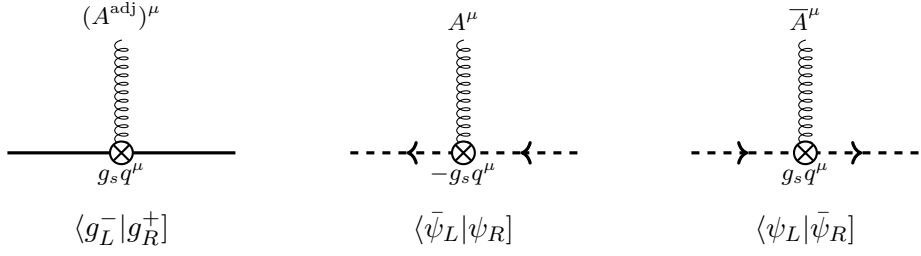

The $\mathcal{O}(g_s)$ order of these three operators are all proportional to $g_sq\cdot \mathbf{A}$, as depicted in Fig.\ref{fig:proportion}.
Boundary contributions corresponding to these boundary operators must be the same except for some color factors. This can be clearly demonstrated in the case of color-ordered amplitudes, which must be proportional to each other,
\begin{equation}\label{boundary-proportional-1}
B^{(3)}_{\langle g^-_L|g^+_R]}~~~\propto~~~ B^{(2)}_{\langle \bar{\psi}_L|\psi_R]}~~~\propto~~~
B^{(0)}_{\langle \psi_L|\bar{\psi}_R]}~.
\end{equation}
It is easy to check that \eqref{boundary-proportional-1} holds for MHV amplitudes. Here we will use the NMHV amplitude as a slightly non-trivial example. The analytic expressions of various NMHV amplitudes can be obtained following \cite{Drummond:2008vq}. Under the $\langle g^-_6|g^+_5]$-shift, the $\mathcal{O}(z^3)$ order boundary contribution of six-gluon NMHV amplitude $A(1^+,2^-,3^+,4^-,5^+,6^-)$ is,
\begin{equation}
B^{(3)}_{\langle g^-_6|g^+_5]}(1^+,2^-,3^+,4^-)
=\frac{\langle 5|2+4|3]^4}
{[23][34]\langle15\rangle\langle56\rangle\langle 1|2+3|4]\langle 5|3+4|2]s_{234}}~.
\end{equation}
Using the amplitude of four gluons and a quark pair, we also verify that,
\begin{equation}
B^{(2)}_{\langle \bar{\psi}_6|\psi_5]}(1^+,2^-,3^+,4^-)
=B^{(0)}_{\langle \psi_6|\bar{\psi}_5]}(1^+,2^-,3^+,4^-)
=-B^{(3)}_{\langle g^-_6|g^+_5]}(1^+,2^-,3^+,4^-)~,
\end{equation}
which is consistent with \eqref{boundary-proportional} and \eqref{boundary-proportional-1}.

%The complete boundary operator can be obtained by translating $g_sq\cdot \mathbf{A}$ from position $x$ to $x-\infty q$ by the action of Wilson line,
%
%\begin{equation}
%\mathbf{W}_q:= \mathbf{W}(x,-\infty q)
%=\mathbf{P}\exp\left(ig\int_{-\infty}^0 \d s~q\cdot A^a(x+sq)\mathbf{T}^a \right)~.
%\end{equation}
%
%In fact, this is a standard way to construct non-local operators [see e.g. 1410.1892]. Suppose $\Phi(x)$ is a quantity whose gauge transformation is
%
%\begin{equation}
%\Phi(x)\rightarrow \Phi(x)\mathbf{V}^{\dagger}(x)~,
%\end{equation}
%
%then $\Phi(x)\mathbf{W}_q$ transforms as
%
%\begin{equation}
%\Phi(x)\mathbf{W}_q\rightarrow \Phi(x)\mathbf{W}_q\mathbf{V}^{\dagger}(-\infty q)~.
%\end{equation}
%
%The $\mathbf{V}^{\dagger}(-\infty q)=1$ if we consider gauge functions which vanish at infinity, therefore  $\Phi(x)\mathbf{W}_q$ is a gauge invariant non-local operator.

%%%%%%%%%%%%%%%%%
\subsection{Wilson line for the gluon-fermion shift}
\label{subsec:g-q-boundary}
%%%%%%%%%%%%%%%%%

%%%%%%%%%%%%%%%%%%%%%%%%%%%%%%%%%%%%%%%%%%%
\subsubsection*{The $\langle \bar{\psi}|g^+]$-shift}
%%%%%%%%%%%%%%%%%%%%%%%%%%%%%%%%%%%%%%%%%%%

In order to obtain $\mathbb{B}^{0}_{21}$, we need to carefully examine the top-left $2\times 2$ block of $\mathbb{V}(\mathcal{D}_0^{-1}\mathbb{V})^k$. A direct computation shows that,
\begin{equation}
\mathbb{B}^{0}_{21}=g_s\sum_{j=0}^{\infty}\sum_{j'=0}^{\infty}
\left( \slashed{A}\frac{ig_s}{2q\cdot\partial} \slashed{q}\right)^j
T^{a_R}\gamma^{\nu}\psi
\left(\frac{ig_s}{q\cdot\partial}q\cdot A^{\tiny\mbox{adj}}\right)^{j'}~,
\end{equation}
in which $q\cdot A^{\tiny\mbox{adj}}=q\cdot A^aT_A^a$ is in the adjoint representation and $\slashed{A}$ is in the fundamental representation. The $q\cdot A^{\tiny\mbox{adj}}$ terms should be regarded as matrix-valued which act on the index $a_R$ in $T^{a_R}$, {\sl i.e.},
\begin{equation}
T^{a_R}\left(\frac{ig_s}{q\cdot\partial}q\cdot A^{\tiny\mbox{adj}}\right)^{k}
:= T^c\Bigl[\left(\frac{ig_s}{q\cdot\partial}q\cdot A^{\tiny\mbox{adj}}\right)^{k}\Bigr]^{c~a_R}~.
\end{equation}
In the large $z$ limit, the external state of hard field $\bar{\psi}(k_L)$ is shifted as $\langle L|\sim z  \langle R|$, and that of $g^+(k_R)$ is shifted as $\epsilon_R^{\mu}(z)\sim \frac{\sqrt{2}z}{\langle LR\rangle}k_L^{\mu}$. Then including the contribution of hard fields, we find the leading large $z$ behavior is $\mathcal{O}(z^2)$, and the boundary operator is
\begin{eqnarray}
\mathcal{B}^{(2)}_{\langle \bar{\psi}_L|g^+_R]}
&=&\frac{\sqrt{2}}{\langle LR\rangle}g_s\sum_{j=0}^{\infty}\sum_{j'=0}^{\infty}
\langle R|\left( \slashed{A}\frac{ig_s}{2q\cdot\partial} \slashed{q}\right)^j
T^{a_R}\slashed{L}\psi
\left(\frac{ig_s}{q\cdot\partial}q\cdot A^{\tiny\mbox{adj}}\right)^{j'}\nonumber\\
&=&-\sqrt{2}g_s\sum_{j=0}^{\infty}\sum_{j'=0}^{\infty}
\left( q\cdot A\frac{ig_s}{q\cdot\partial} \right)^j
T^{a_R}[L|\psi]
\left(\frac{ig_s}{q\cdot\partial}q\cdot A^{\tiny\mbox{adj}}\right)^{j'}~,\label{psibar-g-1}
\end{eqnarray}
%
%This is a Wilson line with a quark insertion.
where in the second line, we have applied the identity \eqref{gamma-change}.

From a first impression of (\ref{psibar-g-1}), one might think $\mathcal{B}^{(2)}_{\langle \bar{\psi}|g^+]}$ is proportional to a product of two Wilson lines, since it depends on two infinite series of operators. However, similar as $\mathcal{B}^{(0)}_{\langle \psi|\bar{\psi}]}$, the $\mathcal{B}^{(2)}_{\langle \bar{\psi}|g^+]}$ can also be simplified using the condition $q\cdot(k_1+\cdots+k_n)=0$, leading to
\begin{equation}
\mathcal{B}^{(2)}_{\langle \bar{\psi}_L|g^+_R]}
=-\sqrt{2}g_sT^{a_R}\sum_{j=0}^{\infty}
\left(q\cdot {A}\frac{ig_s}{q\cdot\partial}\right)^j[L|\psi]
=-\sqrt{2}g_sT^{a_R}W_q^{\dagger}[L|\psi]~,\label{psibar-g-2-0}
\end{equation}
and we will leave the computation detail to appendix \S\ref{appendix:rep}. This means that the boundary contribution of $\langle \bar{\psi}_L|g^+_R]$-shift in Wilson line formalism also follows the same structure as previous cases.
Furthermore, unlike $\mathcal{B}^0_{\langle \psi|\bar{\psi}]}$, the $\mathcal{B}^{(2)}_{\langle \bar{\psi}|g^+]}$ is explicitly gauge invariant even without the condition $q\cdot (k_1+\cdots+k_n)=0$. Therefore there is no need to rewrite it into a $[-\infty,\infty]$ Wilson line as in (\ref{infinite-Wilson}).

%%%%%%%%%%%%%%%%%%%%%%%%%%%%%%%%%%%%%%%%%%%
\subsubsection*{The $\langle \psi|g^+]$-shift}
%%%%%%%%%%%%%%%%%%%%%%%%%%%%%%%%%%%%%%%%%%%

The derivation of $\mathbb{B}^{0}_{31}$ is the same as that of $\mathbb{B}^{0}_{21}$, and we have
\begin{equation}
\mathbb{B}^{0}_{31}=g_s\sum_{j=0}^{\infty}\sum_{j'=0}^{\infty}
\left(  \overline{\slashed{A}}\frac{ig_s}{2q\cdot\partial} \slashed{q}\right)^j
 \overline{T}^b\gamma^{\nu}C^{-1}\bar{\psi}^{\top}
\left(\frac{ig_s}{q\cdot\partial}q\cdot A^{\tiny\mbox{adj}}\right)^{j'}~.
\end{equation}
In the large $z$ limit, the external state of hard fields $\psi_L$ is not shifted, and that of $g^+_R$ is shifted as
\begin{equation}\label{epsilonR}
\epsilon_R^{\mu}(z)= \epsilon_R^{\mu}-z\frac{\sqrt{2}k_L^{\mu}}{\langle LR\rangle}~.
\end{equation}
So at first sight, the large $z$ behavior of boundary contribution is $\mathcal{O}(z)$. However, it turns out the $\mathcal{O}(z)$ boundary operator vanishes,
\begin{eqnarray}
\mathcal{B}^{(1)}_{\langle \psi_L|g^+_R]}
&=&g_s\sum_{j=0}^{\infty}\sum_{j=0}^{\infty}
[L|\left(  \overline{\slashed{A}}\frac{ig_s}{2q\cdot\partial} \slashed{q}\right)^j
 \overline{T}^b\left(\frac{\sqrt{2}}{\langle LR\rangle}\slashed{L}\right)C^{-1}\bar{\psi}^{\top}
\left(\frac{ig_s}{q\cdot\partial}q\cdot A^{\tiny\mbox{adj}}\right)^{j'}\nonumber\\
&=&\frac{\sqrt{2}g_s}{\langle LR\rangle}\sum_{j=0}^{\infty}\sum_{j'=0}^{\infty}
\left(q\cdot  \overline{A}\frac{ig_s}{q\cdot\partial}\right)^j
 \overline{T}^b[L|\slashed{L}|C^{-1}\bar{\psi}^{\top}=0~.\label{psi-g-0}
\end{eqnarray}
The next non-zero contribution is $\mathcal{B}^{(0)}_{\langle \psi_L|g^+_R]}$. To evaluate it, we need firstly to consider the contribution of  $\mathcal{O}(z^0)$ term in \eqref{epsilonR}, which is given by
\begin{eqnarray}
\mathcal{B}^{(0),\I}_{\langle \psi_L|g^+_R]}
&=&g_s\sum_{j=0}^{\infty}\sum_{j'=0}^{\infty}
[L|\left(  \overline{\slashed{A}}\frac{ig_s}{2q\cdot\partial} \slashed{q}\right)^j
 \overline{T}^b\slashed{\epsilon}_RC^{-1}\bar{\psi}^{\top}
\left(\frac{ig_s}{q\cdot\partial}q\cdot A^{\tiny\mbox{adj}}\right)^{j'}\nonumber\\
&=&g_s\sum_{j=0}^{\infty}\sum_{j'=0}^{\infty}
\left(q\cdot  \overline{A}\frac{ig_s}{q\cdot\partial}\right)^j
 \overline{T}^b[L|\slashed{\epsilon}_R|C^{-1}\bar{\psi}^{\top}
\left(\frac{ig_s}{q\cdot\partial}q\cdot A^{\tiny\mbox{adj}}\right)^{j'}\nonumber\\
&=&-\sqrt{2}g_s\frac{[LR]}{\langle LR\rangle}\sum_{j=0}^{\infty}\sum_{j'=0}^{\infty}
\left(q\cdot  \overline{A}\frac{ig_s}{q\cdot\partial}\right)^j
 \overline{T}^b\langle L|C^{-1}\bar{\psi}^{\top}
\left(\frac{ig_s}{q\cdot\partial}q\cdot A^{\tiny\mbox{adj}}\right)^{j'}~,\label{psi-g-1}
\end{eqnarray}
in which we have used
\begin{equation}
[L|\slashed{\epsilon}_R=\frac{[R|\gamma^{\mu}|L\rangle[L|\gamma_{\mu}}{\sqrt{2}\langle RL\rangle}
=-\sqrt{2}\frac{[RL]\langle L|}{\langle RL\rangle}~.
\end{equation}
Similarly, this operator can also be simplified using the condition $q\cdot(k_1+\cdots+k_n)=0$, leading to a boundary contribution that is proportional to a single Wilson line,
\begin{equation}
\mathcal{B}^{(0),\I}_{\langle \psi_L|g^+_R]}
=-\sqrt{2}g_s\frac{[LR]}{\langle LR\rangle}\overline{T}^b\overline{W}_q^{\dagger}
\langle L|C^{-1}\bar{\psi}^{\top}
=-\sqrt{2}g_s\frac{[LR]}{\langle LR\rangle}\langle\bar{\psi}|L\rangle W_qT^b~,\label{psi-g-2}
\end{equation}
where we transposed the whole expression in the last step.

In order to obtain the complete $\mathcal{B}^{(0)}_{\langle \psi_L|g^+_R]}$, we also need to expand $\mathcal{D}_0^{-1}\mathbb{V}$ in \eqref{dv-behaviour} to the next $z$ order. After some tedious computations and simplifications, the final result is\footnote{The infinite sum in \eqref{psi-g-3} cannot be written into a Wilson line, since $q\cdot \partial$ acts on all fields on its right.}
\begin{equation}
\mathcal{B}^{(0),\II}_{\langle \psi_L|g^+_R]}
=-\sqrt{2}g_s\frac{[LR]}{\langle LR\rangle}\langle\bar{\psi}|L\rangle W_qT^{a_L}
+\frac{ig_s}{\sqrt{2}\langle LR\rangle}\bar{\psi}\slashed{L}
\sum_{j=0}^{\infty}\Big(\frac{ig_s}{q\cdot \partial}q\cdot A\Bigr)^{j}
\frac{ig_s}{q\cdot \partial}F_{\alpha\beta}\gamma^{\alpha}\gamma^{\beta}|L]
W_qT^{a_L}~.\label{psi-g-3}
\end{equation}
The complete boundary operator is given by
\begin{equation}
\mathcal{B}^{(0)}_{\langle \psi_L|g^+_R]}
=\mathcal{B}^{(0),\I}_{\langle \psi_L|g^+_R]}
+\mathcal{B}^{(0),\II}_{\langle \psi_L|g^+_R]}~.\label{psi-g-4}
\end{equation}
Similar as \eqref{gluon-not-gauge}, the gauge invariance of this boundary operator requires condition $q\cdot(k_1+\cdots+k_n)=0$. We have computed the form factors
\begin{equation*}
\left\langle0 \Big| \mathcal{B}^{(0)}_{\langle \psi_L|g^+_R]}\Big|g(k_1)\bar{\psi}(k_2)\right\rangle~~~,~~~\left\langle0\Big| \mathcal{B}^{(0)}_{\langle \psi_L|g^+_R]}\Big|g(k_1)g(k_2)\bar{\psi}(k_3)\right\rangle~~~\mbox{and}~~~\left\langle0\Big| \mathcal{B}^{(0)}_{\langle \psi_L|g^+_R]}\Big|g(k_1)g(k_2)g(k_3)\bar{\psi}(k_4)\right\rangle~,
\end{equation*}
and verified that they are in agreement with the boundary contributions of $n$-gluon two-quark amplitudes with $n=2,3,4$ under the $\langle \psi|g^+]$ shift. Some computation details can be found in \S\ref{subsec:g-q-example}.

The expression of $\mathcal{B}_{\langle g^-_{L}|\bar{\psi}_R]}^{(0)}$ is more complicated compared with other boundary operators, since it is essentially a sub-leading large $z$ boundary operator. In general, sub-leading large $z$ boundary operators usually contain multiple Wilson-line-like infinite sums, and are more complicated than the leading order boundary operators.

In practice, the complete boundary operator can be determined by the first a few terms in $g_s$ expansion by gauge invariance, because the higher order terms are present only to make the operator gauge invariant. Therefore, the higher terms can always be organized into Wilson lines, which are natural building blocks of non-local gauge invariant quantities.

%%%%%%%%%%%%%%%
\subsection{Summary of boundary operators in Wilson line formalism}
\label{subsec:summary-boundary}
%%%%%%%%%%%%%%%%

In the previous subsections, we have inspected boundary contributions under various BCFW shifts, and found that all of them contain Wilson lines, which pack infinite series of non-local boundary operator into a simple exponential. The corresponding boundary contributions are therefore related to the form factors of Wilson lines with some field insertions.
% in the Wilson line formalism.  We found that all of them possess the same structure, which is proportional to a single Wilson line with some color and kinematic factors. The Wilson line pack infinite terms of non-local boundary operator into a simple exponential, and the boundary contributions are related computed by the form factor of Wilson line with external non-shifted states.

Summarizing above discussions, we list the Wilson line formalism of boundary contributions in QCD theory as follows for reference. We denote boundary operator as $\mathcal{B}^{\tiny(\mbox{large}~z~\mbox{order}) }_{\tiny\mbox{BCFW~shifts}}$, and ignore the good shifts that produce zero boundary contributions. The boundary operators under $\spab{g^-|g^+}$, $\spab{\bar{\psi}|\psi}$ and $\spab{\psi|\bar{\psi}}$-shifts follow the same expression, except that for two-gluon shift the gauge field and Wilson line are in adjoint representation while the others are in fundamental representation. They are given by,
\begin{equation}
\mathcal{B}_{\langle g^-_{L}|g^+_R]}^{(3)}=g_sq\cdot A^{\tiny\mbox{adj}}~W_q^{\tiny\mbox{adj}}~~~,~~~\mathcal{B}_{\langle \bar{\psi_{L}}|\psi_{R}]}^{(2)}=-g_sq\cdot A~W_q~~~,~~~\mathcal{B}_{\langle \psi_{L}|\bar{\psi}_R]}^{(0)}=-g_sq\cdot \overline{A}~\overline{W}_q~.
\end{equation}
Consequently, the above three boundary operators encode the same Feynman rules except for color factors. In Feynman diagrams of computing boundary contribution, the Wilson line vertices are directly connected to internal or external non-shifted gluons. In case that the configuration of external non-shifted states are the same, the above three boundary operators would also produce the same boundary contribution.

The boundary operators under gluon-fermion shifts are listed as follows,
\begin{eqnarray}
&&\mathcal{B}_{\langle g^-_{L}|\psi_R]}^{(2)}=-\sqrt{2}g_s\langle\bar{\psi}|R\rangle W_qT^{a_L}~~~,~~~
\mathcal{B}_{\langle \bar{\psi}_L|g^+_R]}^{(2)}=-\sqrt{2}g_sT^{a_R}W_q^{\dagger}[L|\psi]~,~~~\label{gm-q-boundary}\\
&&\mathcal{B}^{(0)}_{\langle \psi_L|g^+_R]}
=-\sqrt{2}g_s\frac{[LR]}{\langle LR\rangle}\langle\bar{\psi}|L\rangle W_qT^{a_L}
+\frac{ig_s}{\sqrt{2}\langle LR\rangle}\bar{\psi}\slashed{L}
\sum_{j=0}^{\infty}\Big(\frac{ig_s}{q\cdot \partial}q\cdot A\Bigr)^{j}
\frac{ig_s}{q\cdot \partial}F_{\alpha\beta}\gamma^{\alpha}\gamma^{\beta}|L]
W_qT^{a_L}~,~~~\label{q-gp-boundary}\\
&&\mathcal{B}_{\langle g^-_{L}|\bar{\psi}_R]}^{(0)}=-\sqrt{2}g_s\frac{\langle L R\rangle}{[LR]}T^{a_L}W_q^{\dagger}[R|\psi]
+\frac{ig_s}{\sqrt{2}[ LR]}T^{a_R}W_q^{\dagger}[R|F_{\alpha\beta}\gamma^{\alpha}\gamma^{\beta}\frac{ig_s}{q\cdot \partial}
\sum_{j=0}^{\infty}\Big(\frac{ig_s}{q\cdot \partial}q\cdot A\Bigr)^{j}
\slashed{R}\psi~.~~~\label{gm-qbar-boundary}
\end{eqnarray}
The above boundary operators produce $\mathcal{B}A^n\psi$ or $\mathcal{B}A^n\bar{\psi}$-type Feynman rules with $n\ge 0$.
% gauge fields appear in Wilson line, which after expansion would produce Wilson line vertex coupled to increasing number of gluons. Quark or anti-quark field appears in  the kinematic factor, contributing to a fermion line attached to the Wilson line vertex.
One may also notice that $\mathcal{B}_{\langle g^-_{L}|\psi_R]}^{(2)}$ is proportional to the first term of $\mathcal{B}^{(0)}_{\langle \psi_L|g^+_R]}$ if we replace $|R\rangle\rightarrow |L\rangle$. However, as will be seen in \S\ref{subsec:g-q-example}, their form factors are not related by $k_L\leftrightarrow k_R$, because such a replacement also change the definition of $q$ at the same time.
%From above expressions in the Wilson line formalism, we can clearly see the similarity among boundary operators of different BCFW shifts.

%\textcolor{red}{Comment on why two shifts are different.}

%%%%%%%%%%%%%%%%%%%%%%
\section{Examples of boundary contribution in QCD theory}
\label{sec:example}
%%%%%%%%%%%%%%%%%%%%%%

In this section we will present some examples of QCD amplitudes to demonstrate the computation of boundary contributions in Wilson line formalism.

%%%%%%%%%%%%%%%%%%%%%%%%%%%
\subsection{The BCFW shifts of four-point QCD amplitudes}
%%%%%%%%%%%%%%%%%%%%%%%%%%%

In \S\ref{subsec:4-gluon} we studied the boundary contribution of four-gluon amplitude under $\langle g^-|g^+]$-shift, and in this part we shall study the boundary contributions involving quarks.

%%%%%%%%%%%%%%%%%%%%%%%%%%%%%%%%%%%%
\subsubsection*{Quark-pair $\langle \psi_L|\bar{\psi}_R]$-shift}
%%%%%%%%%%%%%%%%%%%%%%%%%%%%%%%%%%%%%

As discussed in \S\ref{subsec:qq-shift}, the leading large $z$ boundary operator of $\langle \psi_L|\bar{\psi}_R]$-shift is the same as that of $\langle g^-_L|g^+_R]$-shift except for the color structures. Then the two-gluon form factor is given by
\begin{equation}
\Bigl\langle 0\Bigl|~\mathcal{B}_{\langle \psi|\bar{\psi}]}^{(0)}~\Bigr|g^-(k_1)g^+(k_2)\Bigr\rangle=ig_s^2\frac{\langle 1|q|2]^2}{s_{12}\langle 1|q|1]}f^{a_1a_2e}T^e~.~\label{q-qbar-n4}
\end{equation}
The two-gluon two-quark amplitude reads
\begin{equation}
\mathcal{A}(1^-,2^+,3_{\bar{q}},4_q)
=g_s^2\frac{\langle 13\rangle^3\langle 14\rangle}{\langle 12\rangle\langle 23\rangle\langle 34\rangle\langle 41\rangle}T^{a_1}T^{a_2}
+g_s^2\frac{\langle 13\rangle^3\langle 14\rangle}{\langle 21\rangle\langle 13\rangle\langle 34\rangle\langle 42\rangle}T^{a_1}T^{a_2}~,~
\end{equation}
and the boundary contribution under the $\langle \psi_4|\bar{\psi}_3]$-shift is
\begin{equation}
B^{(0)}
=g_s^2\frac{\langle 13\rangle^3}{\langle 12\rangle\langle 34\rangle\langle 23\rangle}[T^{a_1},T^{a_2}]
=ig_s^2\frac{\langle 13\rangle^3}{\langle 12\rangle\langle 34\rangle\langle 23\rangle}f^{a_1a_2 e}T^e~,~
\end{equation}
which is in agreement with \eqref{q-qbar-n4} after setting $|q]=|4], |q\rangle=|3\rangle$.

%%%%%%%%%%%%%%%%%%%%%%%%%%%%%%
\subsubsection*{Gluon-quark $\langle g_L^-|\psi_R]$-shift}
%%%%%%%%%%%%%%%%%%%%%%%%%%%%%%%%

Let us consider the one-gluon one-quark form factor of the boundary operator $\mathcal{B}_{\langle g^-_{L}|\psi_R]}^{(2)}$. The Feynman rules involving this boundary operator are shown in Fig.\ref{fig:rule-gf},
%Feynman rule QCD amplitude gluon fermion shifting
\begin{figure}
\centering
\begin{tikzpicture}
%\draw [help lines, step=0.5] (-1,-3) grid (10,1);
% first rule
  \draw [thick] (0,0)--(1,0);
  \draw [dashed, very thick] (-1,0.05)--(0,0.05);
  \draw [very thick] (-1,-0.05)--(0,-0.05);
  \draw [thick, fill=white] (0,0) circle [radius=0.15];
  \draw [thick] (-0.1,-0.1)--(0.1,0.1) (-0.1,0.1)--(0.1,-0.1);
%  \node [above] at (-1,0) {$a_R$};
  \node [below] at (-1,0) {$a_L$};
%  \node [above] at (-0.5,0.15) {$q$};
%  \draw [->] (-0.7,0.15)--(-0.3,0.15);
  \node [] at (5,0) {$=-\bullet|R\rangle g_sT^{a_L}\sqrt{2}$};
% second rule
  \draw [decorate, decoration={coil, segment length=3}] (0,-2)--(1,-1);
  \draw [thick] (0.5,-2.5)--(1,-3);
  \draw [thick, ->] (0,-2)--(0.5,-2.5);
  \draw [dashed, very thick] (-1,-1.95)--(0,-1.95);
  \draw [very thick] (-1,-2.05)--(0,-2.05);
  \draw [thick, fill=white] (0,-2) circle [radius=0.15];
  \draw [thick] (-0.1,-2.1)--(0.1,-1.9) (-0.1,-1.9)--(0.1,-2.1);
%  \node [above] at (-1,-2) {$a_R$};
  \node [below] at (-1,-2) {$a_L$};
  \node [right] at (1,-1) {$e_1,\mu$};
%  \draw [->] (-0.7,-1.85)--(-0.3,-1.85);
%  \node [above] at (-0.5,-1.85) {$q$};
  \draw [->] (0.9,-1.4)--(0.4,-1.9);
  \draw [->] (0.9,-2.6)--(0.4,-2.1);
  \node [] at (1, -1.7) {$k_1$};
  \node [] at (1, -2.3) {$k_2$};
  \node [] at (5.6,-2) {$=-\bullet|R\rangle g_s^2T^{e_1}T^{a_L}\frac{q^\mu}{q\cdot k_1}\sqrt{2}$};
\end{tikzpicture}
 \caption{Feynman rules of the first and second term of Wilson line operators under gluon-fermion $\spab{g_{L}^-|\psi_{R}}$-shift. The bullet $\bullet$ denotes a contraction with the fermion line emitting from Wilson line operators. $a_L$ is the color index of shifted gluon.}\label{fig:rule-gf}
\end{figure}
and the contributing Feynman diagrams are shown in Fig.\ref{fig:Feynman-4pt-gf}.
%Feynman diagram four-point QCD amplitude gluon fermion shifting
\begin{figure}
\centering
\begin{tikzpicture}
%\draw [help lines, step=0.5] (-1,-1) grid (8,1);
% first diagram
  \draw [dashed, very thick] (-1,0.05)--(0,0.05);
  \draw [very thick] (-1,-0.05)--(0,-0.05);
  \draw [thick] (0,0)--(1,0) (1.5,-0.5)--(2,-1);
  \draw [thick,->] (1,0)--(1.5,-0.5);
  \draw [decorate, decoration={coil, segment length=3}]  (1,0)--(2,1);
  \draw [thick, fill=white] (0,0) circle [radius=0.15];
  \draw [thick] (-0.1,-0.1)--(0.1,0.1) (-0.1,0.1)--(0.1,-0.1);
  \draw [->] (0.3,0.2)--(0.8,0.2);
  \draw [->] (2,0.7)--(1.4,0.1);
  \draw [->] (2,-0.7)--(1.4,-0.1);
%  \node [above] at (-1,0) {$a_R$};
  \node [below] at (-1,0) {$a_L$};
  \node [right] at (2,1) {$\mu, e_1$};
  \node [right] at (2,-1) {};
  \node [] at (0.6,0.4) {$k'$};
  \node [] at (2,0.3) {$k_1$};
  \node [] at (2,-0.3) {$k_2$};
% second diagram
  \draw [very thick, dashed] (5,0.05)--(6,0.05);
  \draw [very thick] (5,-0.05)--(6,-0.05);
  \draw [decorate, decoration={coil, segment length=3}] (6,0)--(7,1);
  \draw [thick, ->] (6,0)--(6.5,-0.5);
  \draw [thick] (6.5,-0.5)--(7,-1);
  \draw [thick, fill=white] (6,0) circle [radius=0.15];
  \draw [thick] (5.9,-0.1)--(6.1,0.1) (5.9,0.1)--(6.1,-0.1);
  \draw [->] (7,0.7)--(6.4,0.1);
  \draw [->] (7,-0.7)--(6.4,-0.1);
%  \node [above] at (5,0) {$a_R$};
  \node [below] at (5,0) {$a_L$};
  \node [right] at (7,1) {$\mu, e_1$};
  \node [right] at (7,-1) {};
  \node [] at (7,0.3) {$k_1$};
  \node [] at (7,-0.3) {$k_2$};
% label
  \node [] at (0.5,-1.5) {$(1)$};
  \node [] at (6,-1.5) {$(2)$};
\end{tikzpicture}
\caption{Two contributing Feynman diagrams for computing boundary contribution of four-point QCD amplitude under $\spab{g_L^-|\psi_R}$-shift.}\label{fig:Feynman-4pt-gf}
\end{figure}
Then we derive
\begin{eqnarray}
\Bigl\langle 0\Bigl|~\mathcal{B}_{\langle g^-_{L}|\psi_R]}^{(2)}~\Bigr|g^+(k_1)\bar{\psi}(k_2)\Bigr\rangle
&=&g_s^2 T^{e_1}T^{a_L}\sqrt{2}\frac{\spaa{2|\slashed{\epsilon}_1\slashed{k}_{12}|R}}{s_{12}}-
g_s^2T^{e_1}T^{a_L}\sqrt{2}\spaa{2~R}\frac{\epsilon_1\cdot q}{q\cdot k_1}\nonumber\\
&=&-2g_s^2 T^{e_1}T^{a_L}\frac{\langle 2R\rangle^2}{\langle 12\rangle\langle 1R\rangle}~.~\label{qbar-q-n4}
\end{eqnarray}
The corresponding amplitude reads
\begin{equation}
\mathcal{A}(1^+,2_{\bar{q}},3^-,4_q)=g_s^2\frac{\langle 32\rangle^3\langle 34\rangle}
{\langle 21\rangle\langle 13\rangle\langle 34\rangle\langle 42\rangle}T^{a_1}T^{a_3}
+g_s^2\frac{\langle 32\rangle^3\langle 34\rangle}
{\langle 23\rangle\langle 31\rangle\langle 14\rangle\langle 42\rangle}T^{a_3}T^{a_1}~.~
\end{equation}
Under the $\langle g_3^-|\psi_4]$-shift, the $\mathcal{O}(z^2)$ boundary contribution is
\begin{equation}
B^{(2)}=-g_s^2\frac{\langle 24\rangle^2}
{\langle 12\rangle\langle 14\rangle}T^{a_1}T^{a_3}~,~
\end{equation}
which is consistent with \eqref{qbar-q-n4} result except for a factor of $2$ which stems from the different definition of $T^a$.

%\todo{A factor of $2$ from kinematic part has been ignored, so the actual computed result is 2 times larger than the compared amplitude result.}

%%%%%%%%%%%%%%%%%%%%%%%%%%%%
\subsection{The gluon-quark shifts of five-point amplitudes}
\label{subsec:g-q-example}
%%%%%%%%%%%%%%%%%%%%%%%%%%%%

In this part we will take a glance at some more complicated examples involving boundary operator form factors with three external states, and using them to illustrate the boundary operators of gluon-quark shifts given by \eqref{gm-q-boundary}-\eqref{gm-qbar-boundary}.

%%%%%%%%%%%%%%%%%%%%%%%%%%%%%%%%%%%g
\subsubsection*{Gluon-quark $\langle g^-_{L}|\psi_R]$-shift}
%%%%%%%%%%%%%%%%%%%%%%%%%%%%%%%%%%

\begin{figure}
  \centering
  \begin{tikzpicture}
%  \draw [help lines, step=0.5] (-1,-2) grid (13.5,6);
% diagram bottow left
    \draw [decorate, decoration={coil, segment length=3}] (0,0)--(2,2) (1,1)--(2,0);
    \draw [thick] (0,0)--(1,-1);
    \draw [dashed, very thick] (-1,0.05)--(0,0.05);
    \draw [very thick] (-1,-0.05)--(0,-0.05);
    \draw [thick, fill=white] (0,0) circle [radius=0.15];
    \draw [thick] (-0.1,-0.1)--(0.1,0.1) (-0.1,0.1)--(0.1,-0.1);
% diagram bottow middle
    \draw [thick] (6,0)--(8,-2);
    \draw [decorate, decoration={coil, segment length=3}] (6,0)--(7,1) (7,-1)--(8,0);
    \draw [dashed, very thick] (5,0.05)--(6,0.05);
    \draw [very thick] (5,-0.05)--(6,-0.05);
    \draw [thick, fill=white] (6,0) circle [radius=0.15];
    \draw [thick] (5.9,-0.1)--(6.1,0.1) (5.9,0.1)--(6.1,-0.1);
% diagram bottow right
    \draw [thick] (12,0)--(13,-1);
    \draw [decorate, decoration={coil, segment length=3}] (12,0)--(13,1) (12,0)--(13.4,0);
    \draw [dashed, very thick] (11,0.05)--(12,0.05);
    \draw [very thick] (11,-0.05)--(12,-0.05);
    \draw [thick, fill=white] (12,0) circle [radius=0.15];
    \draw [thick] (11.9,-0.1)--(12.1,0.1) (11.9,0.1)--(12.1,-0.1);
% diagram top left
    \draw [decorate, decoration={coil, segment length=3}] (1,4)--(3,6) (2,5)--(3,4);
    \draw [thick] (0,4)--(1,4)--(2,3);
    \draw [dashed, very thick] (-1,4.05)--(0,4.05);
    \draw [very thick] (-1,3.95)--(0,3.95);
    \draw [thick, fill=white] (0,4) circle [radius=0.15];
    \draw [thick] (-0.1,3.9)--(0.1,4.1) (-0.1,4.1)--(0.1,3.9);
% diagram top right
    \draw [decorate, decoration={coil, segment length=3}] (7,4)--(8,5) (8,3)--(9,4);
    \draw [thick] (6,4)--(7,4)--(9,2);
    \draw [dashed, very thick] (5,4.05)--(6,4.05);
    \draw [very thick] (5,3.95)--(6,3.95);
    \draw [thick, fill=white] (6,4) circle [radius=0.15];
    \draw [thick] (5.9,3.9)--(6.1,4.1) (5.9,4.1)--(6.1,3.9);
  \end{tikzpicture}
  \caption{Feynam diagrams contributing to the boundary contributions of five-point QCD amplitude under gluon-quark shift. }\label{fig:Feynman-5pt-gf}
\end{figure}
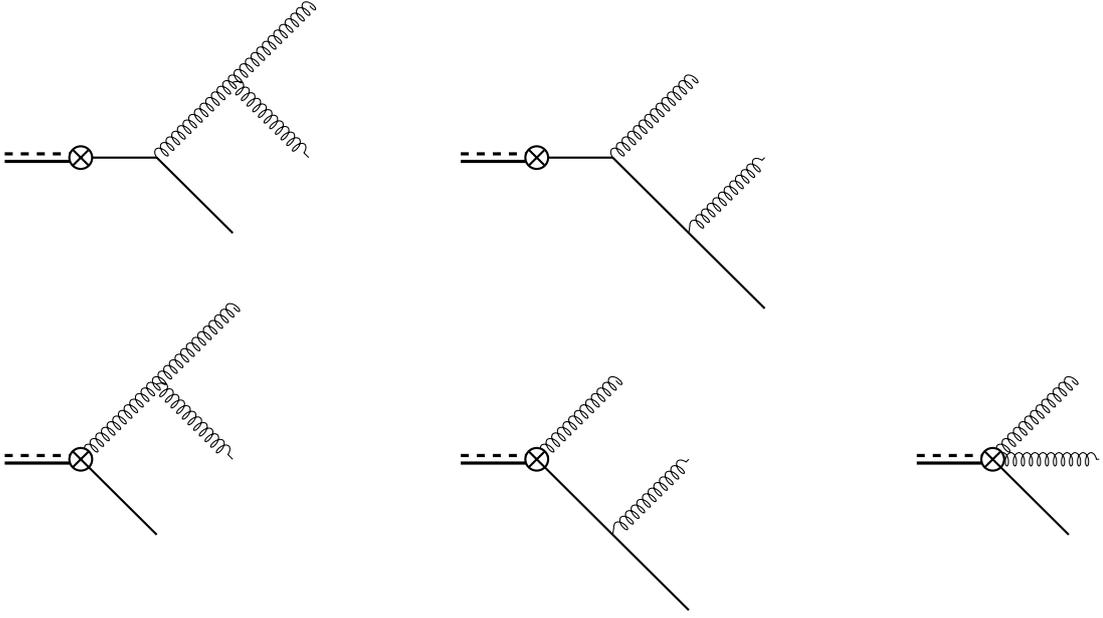

Let us start with the two-gluon one-quark form factor of the boundary operator $\mathcal{B}_{\langle g^-_{L}|\psi_R]}^{(2)}$. The Feynman diagrams are shown in Fig.\ref{fig:Feynman-5pt-gf},
and the Feynman rules can be obtained from the Hermitian conjugate of  \eqref{qbar-gp-feynman-rule}, which is
\begin{equation}
\mathcal{B}A^n\psi~~\rightarrow~~
\bullet|R\rangle g_s^{n+1}q^{\mu_1}\cdots q^{\mu_n}
\frac{T^{a_n}\cdots T^{a_1}T^b}{(q\cdot k_{1\cdots n})\cdots (q\cdot k_{1})}
+\mbox{permutations}~\{1,2,\ldots,n\}~.~\label{gm-q-feynman-rule-R}
\end{equation}
Although the intermediate expressions can be rather lengthy, the final result has a very compact form as,
\begin{equation}
\Bigl\langle 0\Bigl|~\mathcal{B}_{\langle g^-_{L}|\psi_R]}^{(2)}~\Bigr|g^-(k_1)g^+(k_2)\bar{\psi}(k_3)\Bigr\rangle
=-\frac{[25]^3T^{a_2}T^{a_1}T^{a_5}}{[12][35][45][15]}
+\frac{[23][25]^2T^{a_1}T^{a_2}T^{a_5}}{[12][13][35][45]}~.~\label{gm-psi-n5}
\end{equation}

It can be checked that \eqref{gm-psi-n5} produces the correct $\mathcal{O}(z^2)$ boundary contribution of the following amplitude under $\langle g_5^-|\psi_4]$-shift,
\begin{equation}
\mathcal{A}(1^-,2^+,3_{\bar{q}},4_q,5^-)
=\frac{[23][24]^3T^{a_2}T^{a_1}T^{a_5}}{[12][23][34][45][51]}
+\frac{[23][24]^3T^{a_1}T^{a_2}T^{a_5}}{[21][13][34][45][52]}
+\cdots~
\end{equation}
in which we have ignored the partial amplitudes which behaves as $\mathcal{O}(z)$ under the shift.

%%%%%%%%%%%%%%%%%%%%%%%%%%
\subsubsection*{Quark-gluon $\langle \psi_L|g^+_R]$-shift}
%%%%%%%%%%%%%%%%%%%%%%%%%%%%%

Next let us consider the form factor of boundary operator $\mathcal{B}^{(0)}_{\langle \psi_L|g^+_R]}$, with the same external states. This operator has two terms, and the contribution of the first term can be computed similarly as \eqref{gm-psi-n5}. Using the same Feynman diagrams in Fig.\ref{fig:Feynman-5pt-gf}, and similar Feynman rules given as
\begin{equation}
\mathcal{B}A^n\psi~~\rightarrow~~
\bullet|L\rangle g_s^{n+1}\frac{[LR]}{\langle LR\rangle} q^{\mu_1}\cdots q^{\mu_n}
\frac{T^{a_n}\cdots T^{a_1}T^b}{(q\cdot k_{1\cdots n})\cdots (q\cdot k_{1})}
+\mbox{permutations}~\{1,2,\ldots,n\}~.~\label{gm-q-feynman-rule-L}
\end{equation}
we obtain a much more complicated expression compared with \eqref{gm-psi-n5},
\begin{equation}
\begin{aligned}
\Bigl\langle 0\Bigl|\mathcal{B}^{(0),\I}_{\langle \psi_L|g^+_R]}\Bigr|g^-(k_1)&g^+(k_2)\bar{\psi}(k_3)\Bigr\rangle
=-\frac{\langle 13\rangle \langle 25\rangle [25][34]-\langle 14\rangle \langle 15\rangle [14][45]-\langle 15\rangle \langle 34\rangle [34][45]}
{\langle 12\rangle \langle 25\rangle \langle 45\rangle [13][14]}T^{a_1}T^{a_2}T^{a_5}\\
&+\frac{[24]\bigl(\langle 13\rangle \langle 35\rangle [25][34]+\langle 15\rangle \langle 34\rangle [24][45]\bigr)[T^{a_1},T^{a_2}]T^{a_5}}
{\langle 12\rangle \langle 35\rangle \langle 45\rangle [12][14][34]}
+\frac{\langle 13\rangle ^{2}[45]T^{a_2}T^{a_1}T^{a_5}}{\langle 12\rangle \langle 23\rangle \langle 45\rangle [14]}\ .\\
\end{aligned}~\label{gp-psi-n5-1}
\end{equation}

The other contribution of $\mathcal{B}^{(0),\II}_{\langle \psi_L|g^+_R]}$ is relatively easy to compute, which allows us to give more details.
The relevant Feynman rules are
\begin{eqnarray}
&&\mathcal{B}A\psi~\rightarrow \bullet|\slashed{L}\gamma^{\mu_1}\slashed{k}_1|L]\frac{g_s^2T^{a_1}T^{a_R}}{2q\cdot k_1\langle LR\rangle}~,\\
&&\mathcal{B}A^2\psi~\rightarrow\Bigl(
\bullet|\slashed{L}\gamma^{\mu_2}\gamma^{\mu_1}|L]q\cdot k_1
+\bullet|\slashed{L}\gamma^{\mu_2}\slashed{k}_2|L]q^{\mu_1}
+\bullet|\slashed{L}\gamma^{\mu_1}\slashed{k}_1|L]q^{\mu_2}\Bigr)
\frac{g_s^3T^{a_2}T^{a_1}T^{a_R}}{2q\cdot k_1q\cdot k_{12}\langle LR\rangle}+(1\leftrightarrow 2)~.~~~
\end{eqnarray}

Only the last three diagrams in Fig.\ref{fig:Feynman-5pt-gf} contribute, and the result is
\begin{equation}
\begin{aligned}
&\frac{g_s^3[T^{a_1},T^{a_2}]T^{a_R}}{2q\cdot k_{12}s_{12}\langle LR\rangle}
\Big(\epsilon_1\cdot k_2\langle3|\slashed{L}\slashed{\epsilon}_2\slashed{k}_{12}|L]
-\epsilon_2\cdot k_1\langle3|\slashed{L}\slashed{\epsilon}_1\slashed{k}_{12}|L]
+\epsilon_1\cdot \epsilon_2\langle3|\slashed{L}\slashed{k}_{1}\slashed{k}_{2}|L]\Big)\\
&+\frac{g_s^3T^{a_2}T^{a_1}T^{a_R}}{2q\cdot k_1q\cdot k_{12}\langle LR\rangle}
\Big(\langle3|\slashed{L}\slashed{\epsilon}_2\slashed{\epsilon}_1|L]q\cdot k_1
+\langle3|\slashed{L}\slashed{\epsilon}_2\slashed{k}_2|L]q\cdot\epsilon_1
+\langle3|\slashed{L}\slashed{\epsilon}_1\slashed{k}_1|L]q\cdot\epsilon_2\Big)\\
&-\frac{g_s^3T^{a_2}T^{a_1}T^{a_R}}{2q\cdot k_{1}s_{23}\langle LR\rangle}
\langle3|\slashed{\epsilon}_2\slashed{k}_{23}\slashed{L}\slashed{\epsilon}_1\slashed{k}_{1}|L]
+(1\leftrightarrow 2)\ .\\
\end{aligned}\label{q-gp-n5-exp}
\end{equation}
%
%Gauge invariance verified.
The computation can be simplified by setting the reference momenta of gluons to $|r_1]=|L]=|4],|r_2\rangle=|1\rangle$, then
\begin{equation}
\slashed{\epsilon}_1|L]=q\cdot \epsilon_1=k_1\cdot \epsilon_2=\epsilon_1\cdot \epsilon_2=0~,~
\end{equation}
and we find that only two terms in \eqref{q-gp-n5-exp} survive,
\begin{eqnarray}
\Bigl\langle 0\Bigl|\mathcal{B}^{(0),\II}_{\langle \psi_L|g^+_R]}\Bigr|g^-(k_1)g^+(k_2)\bar{\psi}(k_3)\Bigr\rangle
&=&\frac{g_s^3[T^{a_1},T^{a_2}]T^{a_R}}{q\cdot k_{12}s_{12}\langle LR\rangle}
\epsilon_1\cdot k_2\langle3|\slashed{L}\slashed{\epsilon}_2\slashed{k}_{12}|L]
-\frac{g_s^3T^{a_1}T^{a_2}T^{a_R}}{2q\cdot k_{2}s_{13}\langle LR\rangle}
\langle3|\slashed{\epsilon}_1\slashed{k}_{13}\slashed{L}\slashed{\epsilon}_2\slashed{k}_{2}|L] \nonumber \\
&=&-\frac{\langle 34\rangle [24]^{3}[T^{a_1},T^{a_2}]T^{a_5}}{\langle 35\rangle \langle 45\rangle [12][14][34]}
+\frac{[24](s_{14}+s_{34})T^{a_1}T^{a_2}T^{a_5}}{\langle 25\rangle \langle 45\rangle [13][14]}~.~\label{gp-psi-n5-2}
\end{eqnarray}
The sum of \eqref{gp-psi-n5-1} and \eqref{gp-psi-n5-2} produces a simple expression,
\begin{equation}
\Bigl\langle 0\Bigl|~\mathcal{B}^{(0)}_{\langle \psi_L|g^+_R]}~\Bigr|g^-(k_1)g^+(k_2)\bar{\psi}(k_3)\Bigr\rangle
=-\frac{\langle 13\rangle^3 T^{a_2}T^{a_1}T^{a_5}}{ \langle 12\rangle\langle 23\rangle\langle 35\rangle  \langle 45\rangle}
+\frac{\langle 13\rangle^2\langle 15\rangle T^{a_1}T^{a_2}T^{a_5}}
{ \langle 12\rangle \langle 25\rangle\langle 35\rangle\langle 45\rangle}~,~\end{equation}
and it is in agreement with the boundary contribution of the following amplitude under the $\langle \psi_4|g_5^+]$-shift,
\begin{equation}
\mathcal{A}(1^-,2^+,3_{\bar{q}},4_{q},5^+)
=\frac{\langle13\rangle^3\langle14\rangle T^{a_2}T^{a_1}T^{a_5}}{\langle12\rangle\langle23\rangle\langle34\rangle\langle45\rangle\langle51\rangle}
+\frac{\langle13\rangle^3\langle14\rangle T^{a_1}T^{a_2}T^{a_5}}{\langle21\rangle\langle13\rangle\langle34\rangle\langle45\rangle\langle52\rangle}
+\cdots~
\end{equation}
%

%%%%%%%%%%%%%%%%%%%%%%%%%%%%%%%
\subsubsection*{The quark-gluon $\langle \psi_L|g^+_R]$-shift of $\overline{\text{MHV}}$ amplitudes}
%%%%%%%%%%%%%%%%%%%%%%%%%%%%%%%%

We also computed some form factors of $\mathcal{B}^{(0)}_{\langle \psi_L|g^+_R]}$  corresponding to $\overline{\text{MHV}}$ amplitudes, and it turn out that the contribution of $\mathcal{B}^{(0),\II}_{\langle \psi_L|g^+_R]}$ vanishes,
\begin{equation}
\Bigl\langle 0\Bigl|\mathcal{B}^{(0),\II}_{\langle \psi_L|g^+_R]}\Bigr|g^-(k_1)\bar{\psi}(k_2)\Bigr\rangle
=\Bigl\langle 0\Bigl|\mathcal{B}^{(0),\II}_{\langle \psi_L|g^+_R]}\Bigr|g^-(k_1)g^-(k_2)\bar{\psi}(k_3)\Bigr\rangle
=\Bigl\langle 0\Bigl|\mathcal{B}^{(0),\II}_{\langle \psi_L|g^+_R]}\Bigr|g^-(k_1)g^-(k_2)g^-(k_3)\bar{\psi}(k_4)\Bigr\rangle
=0~.~\nonumber %\label{gm-psi-mhvbar}
\end{equation}
For example, it can be easily checked that all terms in \eqref{q-gp-n5-exp} vanish if we set $|r_1]=|r_2]=|L]$. The reason is $\mathcal{B}^{(0),\II}_{\langle \psi_L|g^+_R]}$ contains the factor $F_{\alpha\beta}\gamma^{\alpha\beta}|L]$, in which only the self-dual part of $F_{\alpha\beta}$ survives the projection of $|L]$. Consequently, the form factor vanishes unless there is at least one $g^+$ in the external states.

%We consider the $B(1g^-2g^+3\bar{\psi})$ form factor, which corresponds to the $\langle4|5]$ shift of the MHV amplitude, $\mathcal{A}(1g^-2g^+3\bar{\psi}4\psi5g^+)$.

%%%%%%%%%%%%%%%%%%%%%%
\section{Discussions}
\label{sec:conclu}
%%%%%%%%%%%%%%%%%%%%%%%

In this work, we computed the boundary operators of various BCFW shifts in Yang-Mills theory and QCD, and found that all these boundary operators contain infinite sums which can be conveniently written into infinite or semi-infinite Wilson lines. This Wilson line formalism provides a geometric picture for understanding the infinite series of boundary contributions, and it is applicable to other gauge theories such as $\mathcal{N}=4$ super-Yang-Mills.

From our result we found that, there is a clear resemblance between the boundary operator $\mathcal{B}_{\langle \bar{\psi}_L|g^+_R]}^{(2)}$ in \eqref{gm-q-boundary} and the gauge invariant effective-theory field $\mathcal{X}_{hc}$ \cite{Becher:2006qw},
\begin{equation}
\mathcal{X}_{hc}(x)=\frac{\slashed{n}\slashed{\bar{n}}}{4}W^{\dagger}(x)\psi(x)~,~~~\nonumber
\end{equation}
whose correlation function produces the quark jet function. It is worth to investigate whether the boundary contributions can be used to facilitate the study of quark jet functions, or the other way around.

In \cite{vanHameren:2012uj}, the tree amplitude with one off-shell and $n$ on-shell gluons was decomposed into a gauge invariant part and a gauge restoring part. The gauge invariant part was given by the form factor of the $[-\infty, \infty]$ Wilson line, which is the same as our boundary operator $\mathcal{B}^{(3)}_{\langle g^-_L|g^+_R]}$. Therefore, it is actually equivalent to the boundary contribution of $(n+2)$-gluon amplitudes under $\langle g^-_L|g^+_R]$-shift. Similar relations should also hold for amplitudes with on-shell and off-shell quarks, and in general we think boundary contributions can be regarded as gauge invariant building blocks of off-shell tree amplitudes in QCD and other gauge theories.

%%%%%%%%%%%%%%%%%%%%%%%
\section*{Acknowledgments}
%%%%%%%%%%%%%%%%%%%%

RH is supported by the National Natural Science Foundation of China (NSFC) with Grant No.11805102, Natural Science Foundation of Jiangsu Province with Grant No.BK20180724.

%%%%%%%%%%%%%%%%%%%%%%%%%%
\appendix
%%%%%%%%%%%%%%%%%%%%%

%%%%%%%%%%%%%%%%%%%%%%%%%%%%%
\section{Representations of shifted polarization vectors}
\label{appendix:polarization}
%%%%%%%%%%%%%%%%%%%%%%%%%%%%%%%%%

Here we discuss the polarization vectors of shifted momenta. In $D=4$-dimension, the polarization vectors with momenta $k_L$ and $k_R$ can be expressed in the spinor-helicity form as
\begin{equation}
\epsilon_L^{\mu} |_- = - \frac{\spab{L | \gamma^{\mu} | r_L}}{\sqrt{2}\spbb{r_L~L}}~~~,~~~
\epsilon_L^{\mu}|_+ = - \frac{\spab{r_L | \gamma^{\mu} | L}}{\sqrt{2}\spaa{r_L~L}}~~~,~~~
\epsilon_R^{\mu}|_- = - \frac{\spab{R | \gamma^{\mu} | r_R}}{\sqrt{2}\spbb{r_R~R}}~~~,~~~
\epsilon_R^{\mu}|_+ = - \frac{\spab{r_R | \gamma^{\mu} | R}}{\sqrt{2}\spaa{r_R~R}}~.~
\end{equation}
Taking the following BCFW shift,
\begin{eqnarray}
&&| L \rangle \rightarrow |L \rangle - z | R \rangle ~~~,~~~ |L \rbrack \rightarrow  |L \rbrack~,~~~  \\   &&| R \rangle \rightarrow | R \rangle ~~~,~~~~  |R \rbrack \rightarrow  |R \rbrack + z  |L \rbrack~,~~~
\end{eqnarray}
the polarization vectors become functions of complex variable $z$ as,
\begin{equation}
\begin{aligned}
&\epsilon_L^{\mu} |_-(z) = - \frac{\spab{L | \gamma^{\mu} | r_L}}{\sqrt{2}\spbb{r_L~L}} + z \frac{\spab{R | \gamma^{\mu} | r_L}}{\sqrt{2}\spbb{r_L~L}}  ~~~,~~~~
\epsilon_L^{\mu}|_+(z) = - \frac{\spab{r_L | \gamma^{\mu} | L}}{\sqrt{2}(\spaa{r_L~L}-z\spaa{r_L~R})} ~,~~~   \\
&\epsilon_R^{\mu}|_-(z) = - \frac{\spab{R | \gamma^{\mu} | r_R}}{\sqrt{2}(\spbb{r_R~R}+z \spbb{r_R~L})}  ~~~,~~~
\epsilon_R^{\mu}|_+(z) = - \frac{\spab{r_R | \gamma^{\mu} | R}}{\sqrt{2}\spaa{r_R~R}} - z \frac{\spab{r_R | \gamma^{\mu} | L}}{\sqrt{2}\spaa{r_R~R}}~.~~~
\end{aligned}\label{shiftepsilon}
\end{equation}
If we set the reference momenta as $r_L=k_R$ and $r_R=k_L$, then (\ref{shiftepsilon}) becomes
\begin{equation}\label{newshiftepsilon}
\begin{aligned}
&\epsilon_L^{\mu} |_-(z) = \epsilon_L^{\mu} |_-  - z \frac{\sqrt{2} k_R^{\mu}}{\spbb{L~R}}  ~~~,~~~
\epsilon_L^{\mu}|_+(z) = \epsilon_L^{\mu}|_+ ~,~~~   \\
&\epsilon_R^{\mu}|_-(z) = \epsilon_R^{\mu}|_-  ~~~,~~~
\epsilon_R^{\mu}|_+(z) = \epsilon_R^{\mu}|_+  - z \frac{\sqrt{2} k_L^{\mu}}{\spaa{L~R}}~.~~~
\end{aligned}
\end{equation}
From above result it is simple to get
\begin{equation}
\epsilon_{L}^{\mu} |_-(z)  \cdot  \epsilon_{R}^{\mu}|_+(z)  =  -z^2 + \mathcal{O}(z)~.~
\end{equation}

The above discussions can be generalized to $D$-dimensional space-time, where the shift of polarization vectors follows,
%
%\begin{equation}\label{dshiftepsilon}
%\begin{aligned}
%&\epsilon_L~~\rightarrow ~~\epsilon_L + z\frac{q \cdot \epsilon_L}{k_L\cdot k_R}k_R %~,~~~\\
%&\epsilon_R~~\rightarrow ~~\epsilon_R - z\frac{q \cdot \epsilon_R}{k_L\cdot k_R}k_L~.~~~
%\end{aligned}
%\end{equation}
%
%
\begin{equation}\label{dshiftepsilon}
\epsilon_L~~\rightarrow ~~\epsilon_L + z\frac{q \cdot \epsilon_L}{k_L\cdot k_R}k_R ~~~,~~~\epsilon_R~~\rightarrow ~~\epsilon_R - z\frac{q \cdot \epsilon_R}{k_L\cdot k_R}k_L~,~~~
\end{equation}
where momentum $q=| R \rangle \lbrack L|+|L]\langle R|$.

%%%%%%%%%%%%%%
\section{Reduced Feynman rule of $q\cdot \mathbf{A} \mathbf{W}_q$}
\label{appendix:rep}
%%%%%%%%%%%%%%%%%%%

In \S\ref{subsec:qq-shift} we encounter $q\cdot \mathbf{A}\mathbf{W}_q$ type boundary operators in different representations. We will show that, by applying non-trivial relations of momenta and of color generators, the color structures of Wilson line operators can be simplified, leading to simpler Feynman rules for Wilson line vertices. Furthermore, we show that a Wilson line in an arbitrary representation can be converted to a Wilson line in adjoint representation, which allows us to deal with boundary contributions that involving Wilson lines in different representations.

%%%%%%%%%%%%%%%%%%%%%%%%%%%%%%%%%%%%%%%%%%%
%\subsection*{Reduced Feynman rule of $\langle \psi|\bar{\psi}]$ boundary operator}
%%%%%%%%%%%%%%%%%%%%%%%%%%%%%%%%%%%%%%%%%%%

The Wilson line under two-fermion $\langle \psi|\bar{\psi}]$-shift can be expanded to vertices of Wilson line operators coupled to increasing number of gluons. For vertex of an operator coupled to $n$ gluons, denoted as $\mathcal{B}A^n$-vertex, momentum identity $q\cdot(k_1+k_2\cdots +k_n)=0$ holds. This relation can be used to combine different terms in Feynman rules of vertices.

For the case $n=1$, the $\mathcal{B} A$-vertex can be formally written as,
\begin{equation}
q^{\mu_1}T^{a_1}=q^{\mu_1}\delta^{a_1e_1}T^{e_1}~.
\end{equation}

For the case $n=2$, using $q\cdot k_2=-q\cdot k_1$ the $\mathcal{B} A^2$-vertex can be rewritten as,
\begin{equation}
\frac{q^{\mu_1}q^{\mu_2}}{q\cdot k_2}T^{a_1}T^{a_2}
+\frac{q^{\mu_1}q^{\mu_2}}{q\cdot k_1}T^{a_2}T^{a_1}
=-\frac{q^{\mu_1}q^{\mu_2}}{q\cdot k_1}[T^{a_1},T^{a_2}]
=\frac{q^{\mu_1}q^{\mu_2}}{q\cdot k_1}[T_A^{a_1}]^{a_2e_1}T^{e_1}~,
\end{equation}
where we used $[T_A^a]^{bc}=-if^{abc}$.

Similarly, for the case $n=3$, using $q\cdot k_3=-q\cdot k_2-q\cdot k_1$ the $\mathcal{B} A^3$-vertex can be rewritten as,
\begin{eqnarray}
&&\frac{q^{\mu_1}q^{\mu_2}q^{\mu_3}}{q\cdot k_{23}~q\cdot k_3}
T^{a_1}T^{a_2}T^{a_3}+\text{Permutations}\\
=&&\frac{q^{\mu_1}q^{\mu_2}q^{\mu_3}}{q\cdot k_{23}~q\cdot k_{3}}
[T^{a_1},[T^{a_2},T^{a_3}]]
+\frac{q^{\mu_1}q^{\mu_2}q^{\mu_3}}{q\cdot k_{12}~q\cdot k_2}
[T^{a_2},[T^{a_1},T^{a_3}]]\nonumber\\
=&&\frac{q^{\mu_1}q^{\mu_2}q^{\mu_3}}{q\cdot k_{12}~q\cdot k_{1}}
[T_A^{a_2}T_A^{a_1}]^{a_3e_1}T^{e_1}
+\frac{q^{\mu_1}q^{\mu_2}q^{\mu_3}}{q\cdot k_{12}~q\cdot k_2}
[T_A^{a_1}T_A^{a_2}]^{a_3e_1}T^{e_1}~,~\nonumber
\end{eqnarray}
where we have used the identity,
\begin{equation}
\frac{1}{q\cdot k_{12}~q\cdot k_{2}}
+\frac{1}{q\cdot k_{23}~q\cdot k_{3}}+\frac{1}{q\cdot k_{31}~q\cdot k_{1}}=0~.
\end{equation}
In general, the $\mathcal{B} A^n$-vertex can be reduce to,
\begin{eqnarray}
&&\frac{q^{\mu_1}\cdots q^{\mu_n}}{(q\cdot k_{2\cdots n})\cdots (q\cdot k_n)}T^{a_1}\cdots T^{a_n}+\text{Permutations}~\{1,2,\ldots, n\}\nonumber\\
&=&\frac{q^{\mu_1}\cdots q^{\mu_n}}{(q\cdot k_{1\cdots {n-1}})\cdots (q\cdot k_{n-1})}[T_A^{a_{1}}\cdots T_A^{a_{n-1}}]^{a_ne_1}T^{e_1}+\text{Permutations}~\{1,\cdots,n-1\}~.~\label{wilson-transform}
\end{eqnarray}
The reduced Feynman rule contains $(n-1)!$ terms, while the original Feynman rule contains $n!$ terms. We also notice that in the second line of \eqref{wilson-transform}, all SU($N$) generators are in adjoint representation except for a single overall $T^{e_1}$. This means that a Wilson line in an arbitrary representation can be converted to a Wilson line in the adjoint representation as,
\begin{equation}
q\cdot \mathbf{A} ~\mathbf{W}_q
~\rightarrow~ \sum_{k=0}^{\infty}q\cdot A^a\mathbf{T}^b
\left[\Big(\frac{-g_s}{iq\cdot\partial}q\cdot A^{\tiny\mbox{adj}}\Big)^k\right]^{ab}
=q\cdot A^a
\Big[W_q^{\tiny\mbox{adj}}\Big]^{ab}\mathbf{T}^b~.
\end{equation}
%

%%%%%%%%%%%%%%%%%%%%%%%%%%
\subsection*{The $[-\infty,\infty]$ Wilson line}
%%%%%%%%%%%%%%%%%%%%%%%%%%%%%%

Since the $[-\infty,\infty]$ Wilson lines $\mathbf{W}_q^{[-\infty,\infty]}$ and $q\cdot \mathbf{A} ~\mathbf{W}_q$ are only differed by a $\delta(q\cdot k_{1\cdots m})$ overall factor, following the previous steps, the $[-\infty,\infty]$ Wilson lines can be written as
\begin{equation}
\mathbf{W}_q^{[-\infty,\infty]}=\mathfrak{R}_q^a\mathbf{T}^a~.~
\end{equation}
The quantity $\mathfrak{R}_q^a$ was also discussed in \cite{Kotko:2014aba}, and it can be defined as,
\begin{equation}
\mathfrak{R}_q^a=\frac{1}{T(\mathbf{R})}\tr\Bigl(\mathbf{T}^a\mathbf{W}_q^{[-\infty,\infty]}\Bigr)~,~
\end{equation}
in which $T(\mathbf{R})$ is the index of the representation defined by $\tr(\mathbf{T}^a\mathbf{T}^b)=T(\mathbf{R})\delta^{ab}$. We would like to emphasize that using the reduced Feynman rule in \eqref{wilson-transform}, we have proved that $\mathfrak{R}_q^a$ is independent of the representation.

%%%%%%%%%%%%%%%%%%%%%%%%%%%%%%%%%%%%%%%%%%%
\section{Reduced Feynman rule of gluon-quark shift boundary operator}
\label{appendix:reduce-Feynman-g-q}
%%%%%%%%%%%%%%%%%%%%%%%%%%%%%%%%%%%%%%%%%%%

The boundary operator of gluon-quark shift seems to have more complicated structure. However it can also be reduced by relations of momenta and of color generators, as we would show. For compactness, let us here ignore the $-\sqrt{2}[L|$ factor in \eqref{psibar-g-1}, and focus on the color structures. The Wilson line under gluon-quark shift can be expanded to vertices of operators coupled to one quark and increasing number of gluons, which we denote as $\mathcal{B}A^n\psi$. The Feynman rules can be deduced from these expressions of vertices.

For $\mathcal{B}\psi$-vertex, we simply get the Feynman rule $g_sT^b$. For $\mathcal{B}A\psi$-vertex, using $q\cdot(k_1+k_2)=0$ we get
\begin{equation}
g_s^2\frac{q^{\mu_1}}{q\cdot k_2}T^{a_1}T^b
+g_s^2\frac{q^{\mu_1}}{q\cdot k_1}T^c[T_A^{a_1}]^{cb}
= -g_s^2\frac{q^{\mu_1}}{q\cdot k_1}T^bT^{a_1}~,
\end{equation}
where we have used $T^c[T_A^{a_1}]^{cb}=-if^{a_1cb}T^c=[T^{a_1},T^b]$. For $\mathcal{B}A^2\psi$-vertex, a more tedious computation shows that,
\begin{eqnarray}
&&g_s^3q^{\mu_1}q^{\mu_2}\gamma^{\nu}
\left(\frac{T^{a_1}T^{a_2}T^{b}}{q\cdot k_{23}~q\cdot k_3}
+\frac{T^{a_1}T^{c}[T_A^{a_2}]^{cb}}{q\cdot k_{23}~q\cdot k_2}
+\frac{T^{c}[T_A^{a_1}T_A^{a_2}]^{cb}}{q\cdot k_{12}~q\cdot k_2}
+\Big(1\leftrightarrow 2\Big)\right)\nonumber\\
&=&g_s^3q^{\mu_1}q^{\mu_2}\gamma^{\nu}
\left(\frac{T^{a_1}T^{a_2}T^{b}}{q\cdot k_{23}~q\cdot k_3}
+\frac{T^{a_1}[T^{a_2},T^b]}{q\cdot k_{23}~q\cdot k_2}
+\frac{[T^{a_1},[T^{a_2},T^b]]}{q\cdot k_{12}~q\cdot k_2}
+\Big(1\leftrightarrow 2\Big)\right)\nonumber\\
&=&g_s^3q^{\mu_1}q^{\mu_2}\gamma^{\nu}
\left(\frac{T^bT^{a_1}T^{a_2}}{q\cdot k_{12}~q\cdot k_{1}}
+\frac{T^bT^{a_2}T^{a_1}}{q\cdot k_{12}~q\cdot k_{2}}\right)~,
\end{eqnarray}
where we have used $T^{c}[T_A^{a_1}T_A^{a_2}]^{cb}=-T^{c}f^{a_1ce}f^{a_2eb}
=-i[T^{a_1},T^e]f^{a_2eb}=[T^{a_1},[T^{a_2},T^b]]$. In general, a $\mathcal{B}A^n\psi$-vertex can be rewritten as,
\begin{equation}\label{qbar-gp-feynman-rule}
(-1)^ng_s^{n+1}q^{\mu_1}\cdots q^{\mu_n}
\frac{T^bT^{a_1}\cdots T^{a_n}}{(q\cdot k_{1\cdots n})\cdots (q\cdot k_{1})}
+\mbox{permutations of}~\{1,2,\ldots,n\}~.
\end{equation}
We remark that the permutation is only over gluon legs and this reduced Feynman rule only contain $n!$ terms. The original Feynman rule contains $(n+1)!$ terms and has more complicated color structures.

Using above results, \eqref{psibar-g-1} and \eqref{psi-g-1} can be reduced to the expressions that proportional to a single Wilson line.

\bibliographystyle{JHEP}
\bibliography{wilson}

\end{document}